\documentclass[aps,pre,twocolumn,showpacs,amsmath,amssymb,nofootinbib]{revtex4-1}
\usepackage{default}
\usepackage{amssymb}
\usepackage{color}
\usepackage{amsmath}
\usepackage{mathrsfs}
\usepackage{graphicx}
\usepackage{pdfpages}
\usepackage{subfig}
\usepackage{footnote}

\newcommand{\ch}[1]{\textcolor{black}{#1}}

\begin{document}

\title{Dynamics of localized and patterned structures in the Lugiato-Lefever equation determine the stability and shape of optical frequency combs}
\author{P. Parra-Rivas $^{1,2}$, D. Gomila$^2$, M. A. Mat\'ias$^2$, S. Coen$^3$, L. Gelens$^{1,}$}

\email{lendert.gelens@vub.ac.be}

\affiliation{$^1$Applied Physics Research Group, APHY, Vrije Universiteit Brussel, 1050 Brussels Belgium.\\
$^{2}$Instituto de F\'{\i}sica Interdisciplinar y
  Sistemas Complejos, IFISC (CSIC-UIB),Campus Universitat de les Illes
  Balears, E-07122 Palma de Mallorca, Spain.\\
  $^{3}$Department of Physics,
             The University of Auckland, Private Bag 92019,
            Auckland~1142, New Zealand}
\date{\today}


\begin{abstract}
It has been recently uncovered that coherent structures in microresonators such
as cavity solitons and patterns are intimately related to Kerr frequency combs.
In this work, we present a general analysis of the regions of existence and
stability of cavity solitons and patterns in the Lugiato-Lefever equation, a
mean-field model that finds applications in many different nonlinear optical
cavities. We demonstrate that the rich dynamics and coexistence of multiple
solutions in the Lugiato-Lefever equation are of key importance to understanding
frequency comb generation. A detailed map of how and where to target stable Kerr
frequency combs in the parameter space defined by the frequency detuning and the
pump power is provided. Moreover, the work presented also includes the
organization of various dynamical regimes in terms of  bifurcation points of
higher co-dimension in regions of parameter space that were previously
unexplored in the Lugiato-Lefever equation. We discuss different dynamical
instabilities such  as oscillations and chaotic regimes.
\end{abstract}

\maketitle
\section{Introduction}

\begin{figure}
\begin{center}
 \includegraphics[width = 8cm]{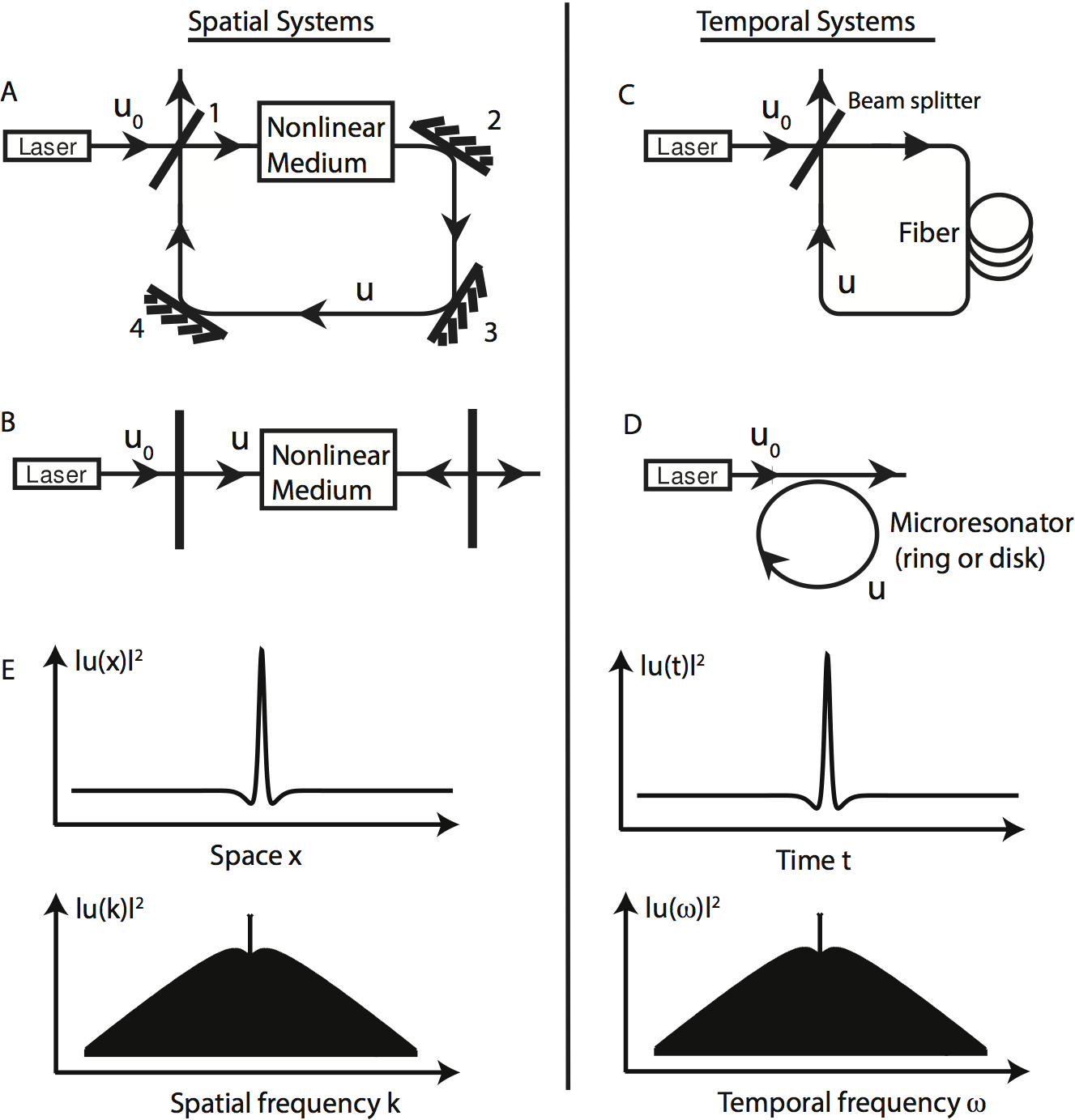}
\end{center}
\caption{Different set-ups where the mean field $u$ is described by the
Lugiato-Lefever equation (LLE). A. A ring cavity partially filled with a
nonlinear medium. Mirror 1 partially transmits the input beam $u_{0}$, while
mirrors 2-4 are completely reflecting. B. A Fabry-Perot resonator filled with a
nonlinear medium. C. A nonlinear all-fiber cavity. D. A microresonator where the
cavity can be either a ring or a disk. E. A sketch of a localized structure,
both in the time (spatial) domain and in the corresponding Fourier domain.}
\label{Fig::LLE_setups}
\end{figure}

Optical frequency combs can be used to measure light frequencies and time intervals more easily and precisely than
ever before \cite{SciAmFreqCombs}, opening a large avenue for applications. \ch{Traditional} frequency \ch{combs are
usually associated with trains} of evenly spaced, very short pulses. More recently, a new generation of \ch{comb
sources} has been \ch{demonstrated} in compact \ch{high-Q} optical microresonators \ch{with a Kerr nonlinearity
pumped by continuous-wave laser light} \cite{delhaye_optical_2007}. \ch{These combs are now referred to} as Kerr
frequency combs and \ch{have attracted a lot of interest in the last few years \cite{Kippenberg_science_2011}}.
Interestingly, it has been demonstrated that Kerr frequency combs can be modeled in a way that is strongly
reminiscent of \ch{temporal} cavity solitons (CSs) in nonlinear cavities
\cite{coen_modeling_2012,coen_scalinglaws_2013,Chembo_LLE_2013}. \ch{Temporal CSs have been experimentally studied
in fiber resonators \cite{leo_temporal_2010} and their} description is based on a now classical equation, the
Lugiato-Lefever equation (LLE), that describes pattern formation in optical systems \cite{LL-1987}.

Lugiato and Lefever\ch{, in 1987,} introduced \ch{the} mean-field model \ch{of} nonlinear optical cavities, in which
alternation of propagation around the cavity with coherent addition of an input field is replaced by a single
partial differential equation with a driving term \cite{LL-1987}. The LLE is applicable
\ch{to} different types of cavities, as shown in Figure\ \ref{Fig::LLE_setups}. It was originally derived to
describe a ring cavity \ch{or} a Fabry-Perot resonator \ch{with a transverse spatial extension and} partially filled
with a nonlinear medium \cite{LL-1987} (see Figures\ \ref{Fig::LLE_setups}A-B). The LLE can also be used in the
context of \ch{single-mode} fiber cavities (see Figure\ \ref{Fig::LLE_setups}C) \cite{leo_temporal_2010}. In this
case, the spatial coordinate in the LLE for a spatially extended cavity with diffraction is replaced by a time
coordinate to model chromatic dispersion of light \ch{in the longitudinal (temporal) direction}. Finally, more
recently, the LLE has been \ch{applied to} microresonators (see Figure\ \ref{Fig::LLE_setups}D) in the context of
optical frequency combs
\ch{\cite{matsko_mode-locked_2011,coen_modeling_2012,coen_scalinglaws_2013,Chembo_LLE_2013}}.


The interplay between diffraction and/or dispersion and nonlinearity \ch{can lead} to \ch{the formation of} complex
spatio-temporal structures inside the cavity, such as patterned and localized solutions (see Figure\
\ref{Fig::LLE_setups}E). In the cavities presented in Figures \ref{Fig::LLE_setups}A-B, a spatially localized bright
light spot embedded in a \ch{homogeneous} background of light has been shown to exist at the output of the resonator
\cite{2002Natur.419..699B}. Such structures, localized in space, are also known as \textit{spatial} cavity solitons
(CSs). Similarly, in the cavities presented in Figures \ref{Fig::LLE_setups}C-D, a stable structure that is
localized in time can exist, also known as \textit{temporal} cavity solitons \cite{leo_temporal_2010}. A sketch of
such a localized structure, both in the time (spatial) domain \ch{as well as} in the \ch{corresponding Fourier}
domain, is shown in Figure\ \ref{Fig::LLE_setups}E.

The formation and  stability of patterns and CSs have been theoretically studied in great detail in the LLE, both in
a one dimensional (1D) and a two dimensional (2D) setting
\cite{LL-1987,mcsloy_computationally_2002,Scroggie,Gomila_PhysD, 2002JOSAB..19..747F, Damia_PRE}. The theory
\ch{developed} for CSs in the LLE was mainly motivated by the potential application of spatial CSs as information
carriers in all-optical memories~\cite{2002Natur.419..699B,mcdonald_spatial_1990}. \ch{Recent} experimental
observations of 1D temporal CSs \ch{in fiber resonators \cite{leo_temporal_2010} have renewed the interest in the
LLE. This interest was further strengthened when it was demonstrated that the LLE can also be used to efficiently
model Kerr frequency combs and that these were found, in some conditions, to be} closely related to temporal CSs
present inside the cavity \cite{coen_modeling_2012,coen_scalinglaws_2013,Chembo_LLE_2013}. \ch{Experimental evidence
of CSs in microresonators has emerged shortly after \cite{Herr_2013}}. The theory that has been developed for
spatial CSs can thus potentially provide important information about the properties of Kerr frequency combs. In 1D,
CSs have mainly been studied for lower values of the cavity frequency detuning, where \ch{they} were shown to be
always stable \cite{Gomila_PhysD}. However, in the context of Kerr frequency combs, experiments are often carried
out at \ch{relatively high pump power \cite{delhaye_optical_2007,foster_silicon-based_2011}. Because of the
Kerr-induced tilt of the cavity resonance, the detuning, which is measured from the center of the linear (cold)
resonance, is correspondingly large. Only a few recent studies have been reported in this regime
\cite{coen_scalinglaws_2013,Vladimirov_2012,Leo_Gelens_2013}. In particular, a fiber cavity experiment has
demonstrated that CSs can} display periodic oscillations, and preliminary numerical analysis have also revealed
various chaotic regimes \cite{Leo_Gelens_2013}.


In this work, we aim \ch{to realize} two goals: i) to interpret how various coherent structures in microresonators,
such as patterns and solitons, are intimately related to different types of Kerr frequency combs; ii) to expand the
study of the LLE to operating regimes that will prove to be of key importance for frequency comb generation, but so
far have not been much explored. Such a detailed analysis of the unfolding of the rich \ch{dynamical behavior of}
the LLE for higher values of the detuning will also prove to be of fundamental interest for researchers working in
the field of dissipative solitons.

The organization of this paper will be as follows. In Section \ref{Sec::LLE}, we will introduce the Lugiato-Lefever
model in more detail. In Section \ref{Sec::snaking} we will discuss the coexistence of multiple solutions, such as
homogeneous solutions, patterned solutions and localized solutions. We will show that all these structures are
organized in a so-called \textit{snaking} diagram and will \ch{discuss} its importance in \ch{the context of}
frequency comb generation. \ch{In Section \ref{Sec::spatdyn}, we consider} how the homogeneous state can connect to
a patterned state and back, thus forming CSs. \ch{For transverse systems, this is referred to as the \emph{spatial}
dynamics and we will keep this terminology for combs for which the spatial coordinate is replaced by a longitudinal
(temporal) variable.} This study provides us with essential information about the location in parameter space where
stable CSs and frequency combs can be found. In Section \ref{Sec::dyn1CS} the dynamical behavior of single-peak CSs
and their corresponding frequency combs is studied in the parameter space defined by the cavity detuning and the
pump power. Finally, we \ch{draw our conclusions} in the final Section \ref{Sec::discussion}.

\section{The Lugiato-Lefever equation}\label{Sec::LLE}

The LLE is a prototype model describing an optical cavity filled with a nonlinear Kerr
medium as derived by Lugiato and Lefever in order to study pattern formation in this system \cite{LL-1987}. Later
studies have then demonstrated the existence of CSs in the LLE
\cite{mcsloy_computationally_2002,Scroggie,Gomila_PhysD,2002JOSAB..19..747F, Damia_PRE}. The LLE was originally
obtained through a mean-field approximation, describing the dynamics of the slowly varying amplitude of the
electromagnetic field $u(\varphi,t)$ in the paraxial limit, where $\varphi=x$ is the spatial coordinate \ch{transverse} to the
propagation direction. Here we consider only one transverse dimension. The time evolution of the electric field
\ch{over many cavity roundtrips} can then be described as follows after a suitable rescaling of the variables
\cite{LL-1987}:
\begin{equation}\label{eq.1}
\partial_{t}u=-(1+i\theta)u+i|u|^{2}u+u_{0}+i \partial_{\varphi}^{2} u\,.
\end{equation}
The first term on the right-hand side describes cavity losses (\ch{the system is dissipative by} nature); $u_0$ is
\ch{the amplitude of} the homogeneous (plane wave) input field or pump; $\theta$ \ch{measures} the cavity frequency
detuning between the frequency of the input pump and the nearest cavity resonance; $
\partial_{\varphi}^{2}$ models diffraction; and the sign of the cubic term is \ch{set so} that it corresponds to the
self-focusing case. In the case of a fiber resonator \ch{or a whispering-gallery-mode (WGM) microresonator}, the LLE
still holds, but \ch{the input field $u_0$ is to be interpreted as the amplitude of the continuous-wave pump beam}
while the field $u(\varphi,t)$ is now function of two time scales, with $\varphi=\tau$ a time coordinate \ch{--- also known as
 \emph{fast} time ---} in the frame moving with the group velocity
\cite{leo_temporal_2010,coen_modeling_2012,coen_scalinglaws_2013}. The spatial derivatives are thus replaced by time
derivatives $\partial_{\varphi}^{2} u =
\partial_{\tau}^{2} u$, \ch{and} the sign of \ch{this} term is such that
one has anomalous dispersion. \ch{Note that in our study here, we only consider second-order dispersion. Several works have also studied the influence of higher order dispersion on the dynamics of dissipative structures in the LLE\cite{Gelens_PRA_2008,Leo_prl_2013,GelensOL2010}.} In the
case of \ch{WGM} microresonators, the LLE has also been shown to remain valid when replacing the coordinate $\varphi$ by
the resonator's azimuthal angle \cite{Chembo_LLE_2013}.

\begin{figure*}[t!]
\centering
\includegraphics[width=2 \columnwidth]{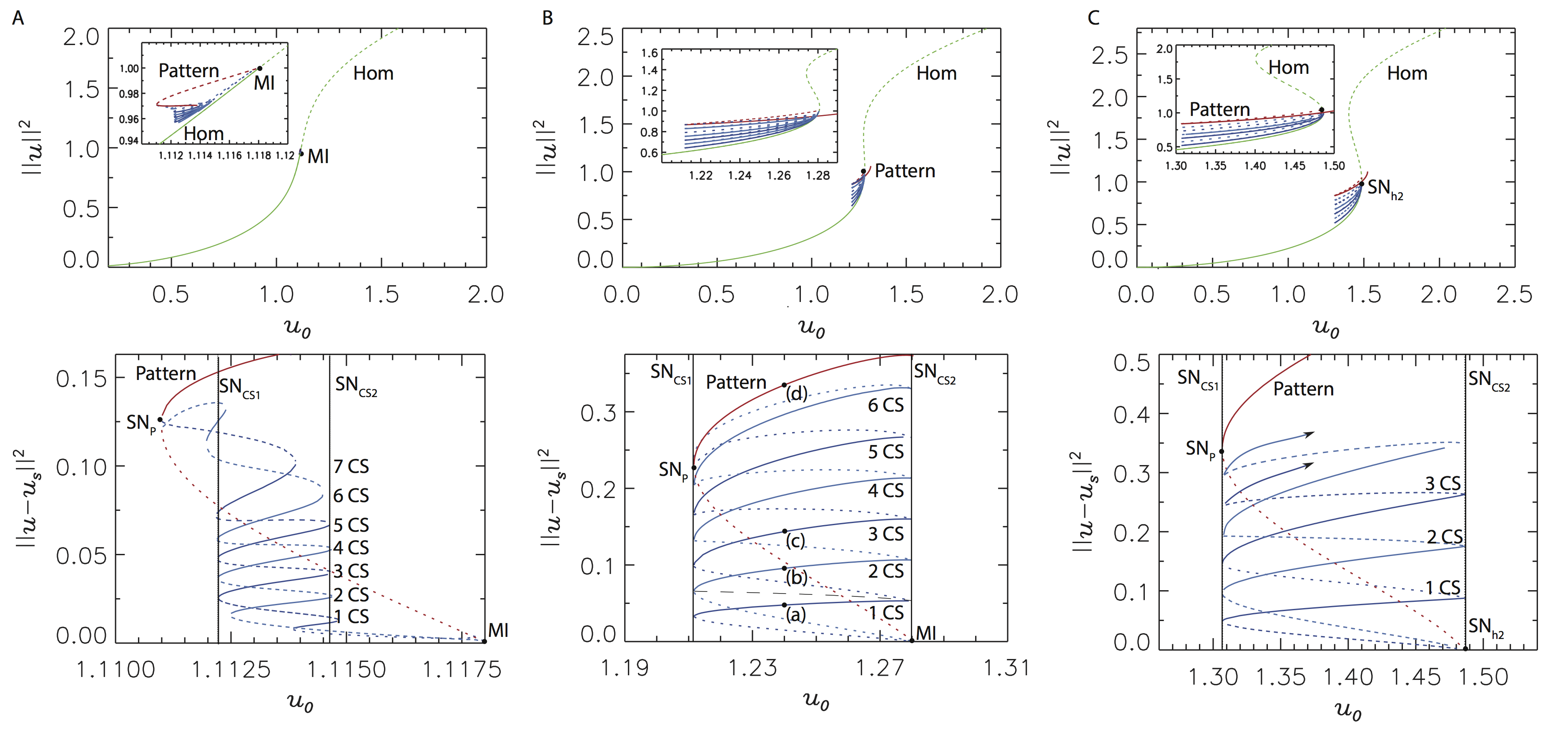}
\caption{(Color online) Various solution branches of the LLE as function of
the pump $u_0$ for different values of the detuning $\theta$ [$\theta = 1.5$
(A), $\theta = 1.8$ (B),
$\theta = 2.1$ (C)]. The (green) line denoted by ``Hom'' shows the homogeneous
steady states (HSS) solutions \ch{of} Eq.\ (\ref{eq.1}), where the inset shows a zoom of
the region around the modulational instability (MI). A patterned branch (red
curve, denoted ``Pattern'') is created subcritically in MI. The bottom panels show
a more detailed picture of the snaking region where CSs \ch{(blue curve)} exist, plotting the
energy of the solution after removal of the homogeneous background. The snaking
region is located between two other saddle-node bifurcations, called
$\mathrm{SN}_{\mathrm{CS}1,2}$. All solid (dotted) lines correspond to stable (unstable)
solutions. The dots denoted by (a)--(d) in panel B correspond to solutions of
which the profile is shown in Figure \ref{profiles}.}
\label{fig::Hom_plus_snaking}
\end{figure*}

\begin{figure*}[t!]
\centering
\includegraphics[width=2 \columnwidth]{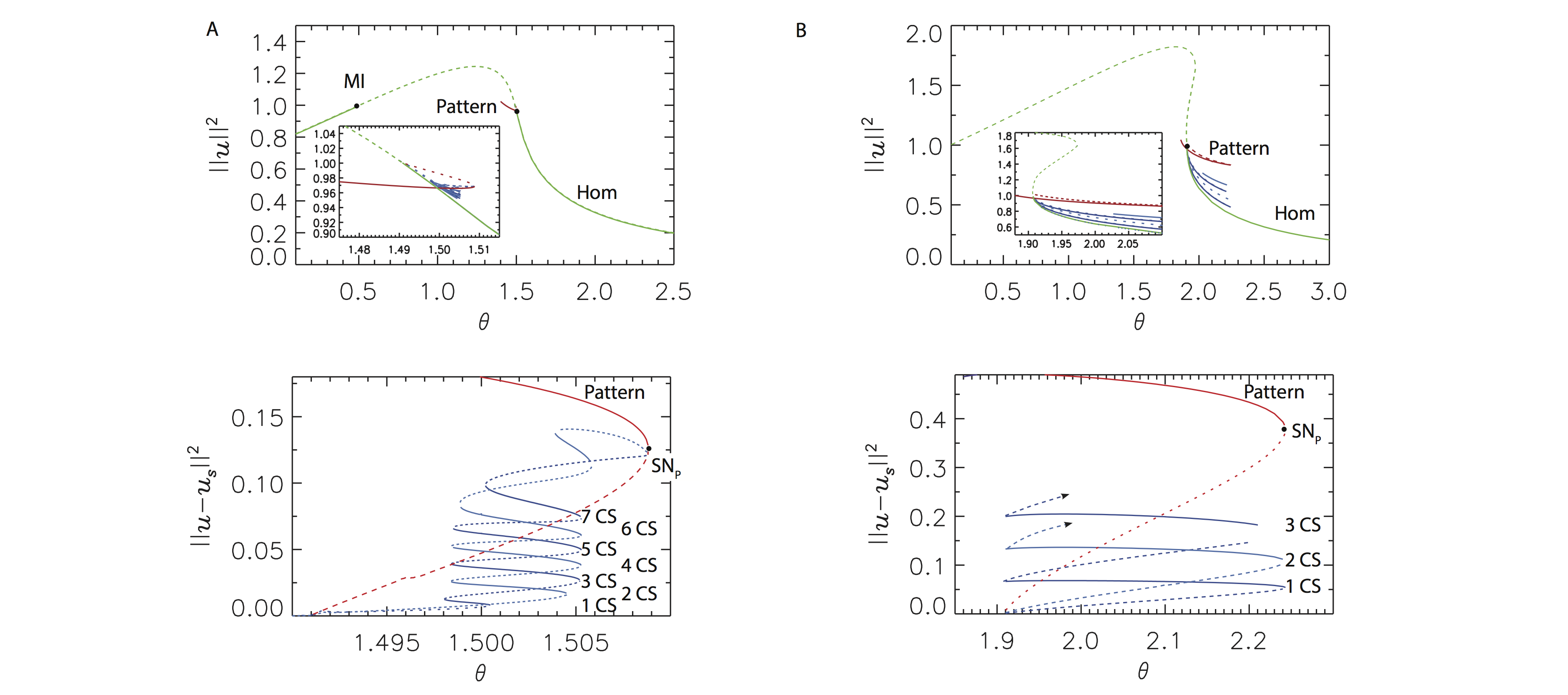}
\caption{(Color online) \ch{Same as Figure~\ref{fig::Hom_plus_snaking} but} as function of
the detuning $\theta$ for different values of the pump $u_0$ [$u_0 =
1.114$ (A), $u_0 = 1.35$ (B)]. }
\label{fig::Hom_plus_snaking_det}
\end{figure*}

The homogeneous steady state (HSS) solutions $u_{s}$ of the LLE (\ref{eq.1}) are easily found by \ch{setting} all
derivatives to be zero:
\begin{equation}
	I_s [1+(\theta - I_s)^{2}] = I_0\,.
	\label{eq:StationarySolutionLLE}
\end{equation}
\ch{The above equation is the classic cubic equation (S-shape response) of dispersive optical bistability}, with
$I_s = |u_{s}|^{2}$ and $I_0 = |u_{0}|^{2}$. For $\theta < \sqrt{3}$, \ch{only one homogeneous solution exists,
hence the system is monostable. In the context of spatial CSs, this monostable regime has been studied in great
detail \cite{LL-1987,mcsloy_computationally_2002,Scroggie,Gomila_PhysD, 2002JOSAB..19..747F, Damia_PRE}. Much less
analysis has been done \ch{when} $\theta > \sqrt{3}$. \ch{In this case, for a range of input intensities $I_0$
Eq.~(\ref{eq:StationarySolutionLLE}) has three homogeneous solutions, out of which one is a saddle solution. Hence the
system is often called to be bistable. In this bistable region} much more complex dynamics are observed in the LLE
\cite{Leo_Gelens_2013}.}

One can easily show the existence of two HSSs analytically \ch{by looking for the points where} the derivative
$\partial I_0/\partial I_s$ is equal to zero,
\begin{equation}
	\frac{\partial I_0}{\partial I_s} = 1 + (\theta - I_s)^{2} - 2(\theta - I_s)I_s = 0 \text{ .}
	\label{eq:Slope_LLE}
\end{equation}
\ch{The solutions to this equation give the turning points of the bistable response, also known as} the saddle-node
\ch{(SN)} bifurcations of the HSSs:
\begin{equation} I_s^{\mathrm{SN}_{h1,2}}
	= \frac{2 \theta}{3} \pm \frac{1}{3} \sqrt{\theta^{2}-3} \text{ .}
	\label{eq:turningPoints}
\end{equation}
It is clear that for $\theta^{2} < 3$ there are no turning points (see Figure \ref{fig::Hom_plus_snaking}A), while
for $\theta^{2} > 3$ there are two, $\mathrm{SN}_{h1,2}$, and the system is {\it bistable} (see Figures
\ref{fig::Hom_plus_snaking}B-C). Figures \ref{fig::Hom_plus_snaking}A-C are plotted versus the input power \ch{for
selected detunings while Figs. \ref{fig::Hom_plus_snaking_det}A-B presents similar plots but as a function of
detuning for selected pump powers. This second set of plots may be more relevant to the microresonator Kerr
frequency comb context, in which the pump laser frequency is used as an important control parameter and for which
thermal effects also strongly affect the detuning.}

Analyzing the linear stability of the HSS to perturbations of the form $\exp{(i k \varphi
 + \sigma t)}$ \ch{\cite{Scroggie}}, one finds that, for $\theta < \sqrt{3}$, the HSS loses its
stability at $I_s =1$ with critical wavenumber $k_c=\sqrt{2-\theta}$ at which point a patterned solution is created
either supercritically ($\theta < 41/30$) or subcritically ($\theta > 41/30$) \cite{Scroggie}. In the bistable regime ($\theta >
\sqrt{3}$), one can differentiate two different cases: $\sqrt{3} < \theta < 2$, in which, as before, the MI occurs
at $I_s=1$, and $\theta > 2$ in which the critical wavenumber is zero and the HSS is stable all the way to
$\mathrm{SN}_{h2}$. In the latter case if $I_s$ is increased above $I_s^{\mathrm{SN}_{h2}}$ the HSS jumps to the
upper branch, but the upper branch HSS has always a wide range of unstable wavenumbers, whose growth leads to a
spatio-temporal chaotic regime called optical turbulence \cite{CH,Gomila03}.

In this work we will show how localized and patterned solutions are organized in both the monostable and the
bistable regime, putting more emphasis on the bistable region as the monostable one has already been studied in
depth. A typical  example of the bifurcation diagram of the homogeneous solutions is shown in Figures
\ref{fig::Hom_plus_snaking}A-C for values of the detuning $\theta$ in each characteristic region: $\theta <
\sqrt{3}$, $\sqrt{3} < \theta < 2$, and $\theta > 2$. In the bistable region shown in Figures
\ref{fig::Hom_plus_snaking}B-C, there exists a small region of input powers where \ch{the} two homogeneous solutions
coexist. The homogeneous solution with the highest intensity is always unstable in Eq.\ (\ref{eq.1}). A pattern is
created subcritically at the modulational instability for $\theta<2$ such that a stable pattern can coexist with the
stable homogeneous solution in a certain range of parameters. For  $\theta>2$, although the system goes directly to
optical turbulence above $\mathrm{SN}_{h,2}$, stable periodic patterns persist below threshold. Such coexistence is
known to potentially give rise to localized structures, such as CSs, as will be discussed in the next Section.

\section{Cavity solitons, patterns and frequency combs}\label{Sec::snaking}

The bottom panels in Figures \ref{fig::Hom_plus_snaking}A-C and \ref{fig::Hom_plus_snaking_det}A-B show a zoom of
the region where a subcritical branch of spatially periodic states \ch{(red)} coexists with a stable homogeneous
solution \ch{(green)}. For $\theta> 41/30$ the pattern branch is unstable at its point of origin (MI), but acquires
stability at finite amplitude at a saddle-node bifurcation $\mathrm{SN}_P$. In addition, it is known that there are
two branches of CSs \ch{(blue)} that bifurcate from the homogeneous solution simultaneously with the periodic
states, and do so likewise subcritically \cite{Gomila_PhysD}. These states are therefore also initially unstable.
When followed numerically these states become better and better localized and once their amplitude and width become
comparable to the amplitude and wavelength of the periodic state, these CS states begin to grow in spatial extent by
adding peaks symmetrically \ch{on} either side\ch{, thereby forming bound states of CSs}. In a bifurcation diagram
this growth is associated with back and forth oscillations across a pinning interval
[$\mathrm{SN}_{\mathrm{CS}1}-\mathrm{SN}_{\mathrm{CS}2}$]. This behavior is known as \textit{homoclinic snaking}
\cite{Champneys98,Champneys00,KnoblochJ,Burke06,Kozyreff-2006}, and is associated with repeated gain and loss of
stability of the associated localized structures. This snaking structure has been experimentally verified for
spatial CSs in a semiconductor-based optical system \cite{Barbay_snaking}.

Figure \ref{fig::Hom_plus_snaking}  shows a clear difference in the location of the pinning interval
[$\mathrm{SN}_{\mathrm{CS}1}-\mathrm{SN}_{\mathrm{CS}2}$] when transitioning from the monostable (Figure
\ref{fig::Hom_plus_snaking}A) to the bistable region of operation (Figures \ref{fig::Hom_plus_snaking}B-C). For
$\theta < \sqrt{3}$, the pinning interval where CSs can be found \ch{spans only a part} of the region of coexistence
between the stable pattern and the stable homogeneous solution, while for increasing values of the detuning in the
bistable region ($\theta > \sqrt{3}$) the pinning interval quickly spans this entire region of coexistence. Another
clear trend is that the CSs exist over a much broader range of values of the input \ch{amplitude $u_0$ and detuning}
$\theta$ as both $u_0$ and $\theta$ are increased in the bistable region. This evolution will be discussed in more
detail in the next Section \ref{Sec::spatdyn}. We remark that for increasing values of the detuning $\theta$ it
becomes increasingly difficult to numerically track the CS snaking branches. We believe this is not a numerical
artifact, but rather that the snaking structure becomes more complex for $\theta >2$ hindering the numerical
tracking of these CS branches (although solutions with multiple CSs seem to persist). Understanding the detailed
snaking structure for high values of $\theta$ is beyond the scope of this paper and it will addressed elsewhere.

\begin{figure}[t!]
\centering
\includegraphics[scale=0.5]{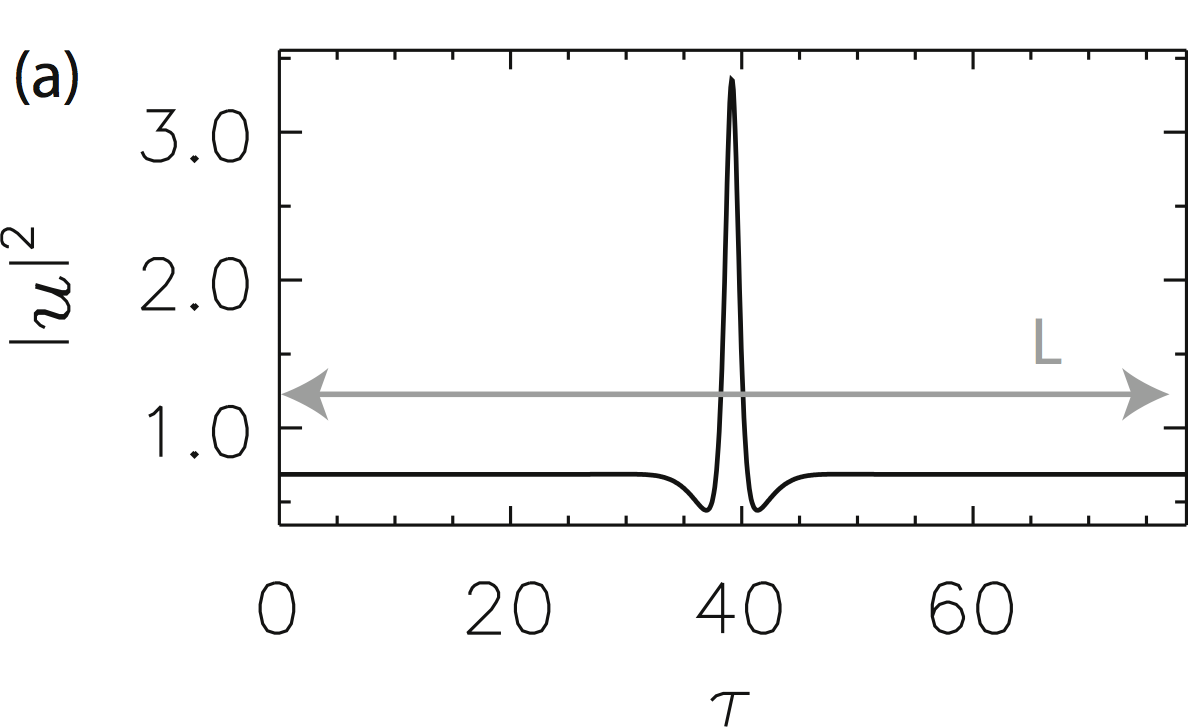}
\includegraphics[scale=0.5]{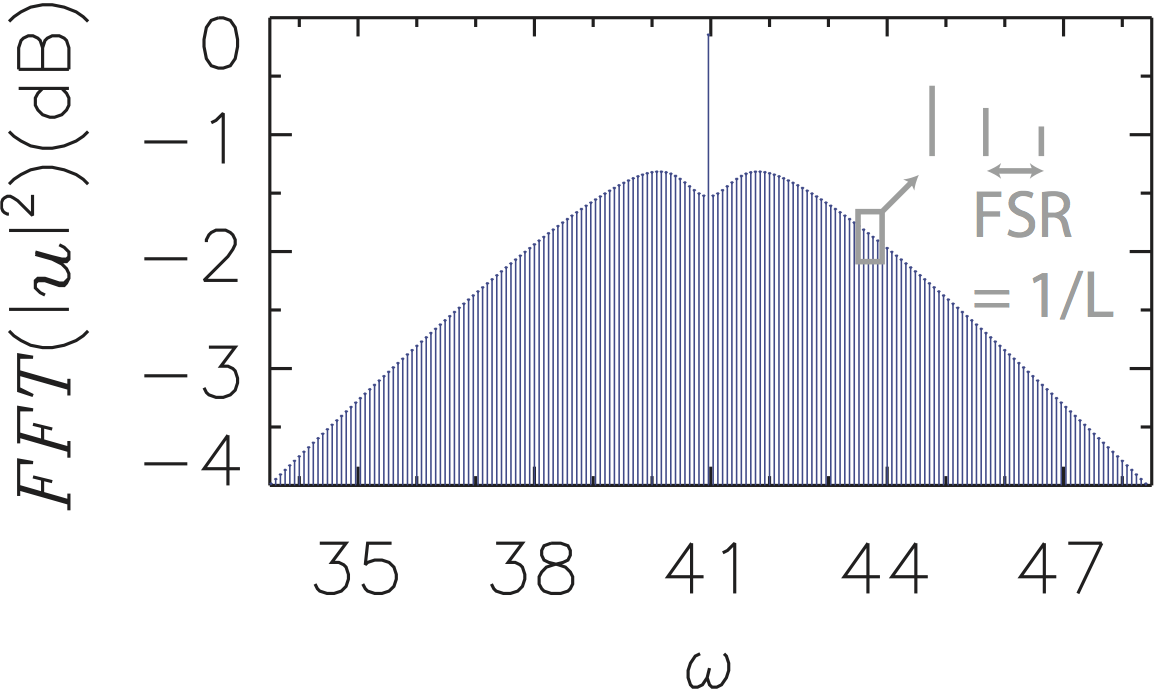}
\includegraphics[scale=0.5]{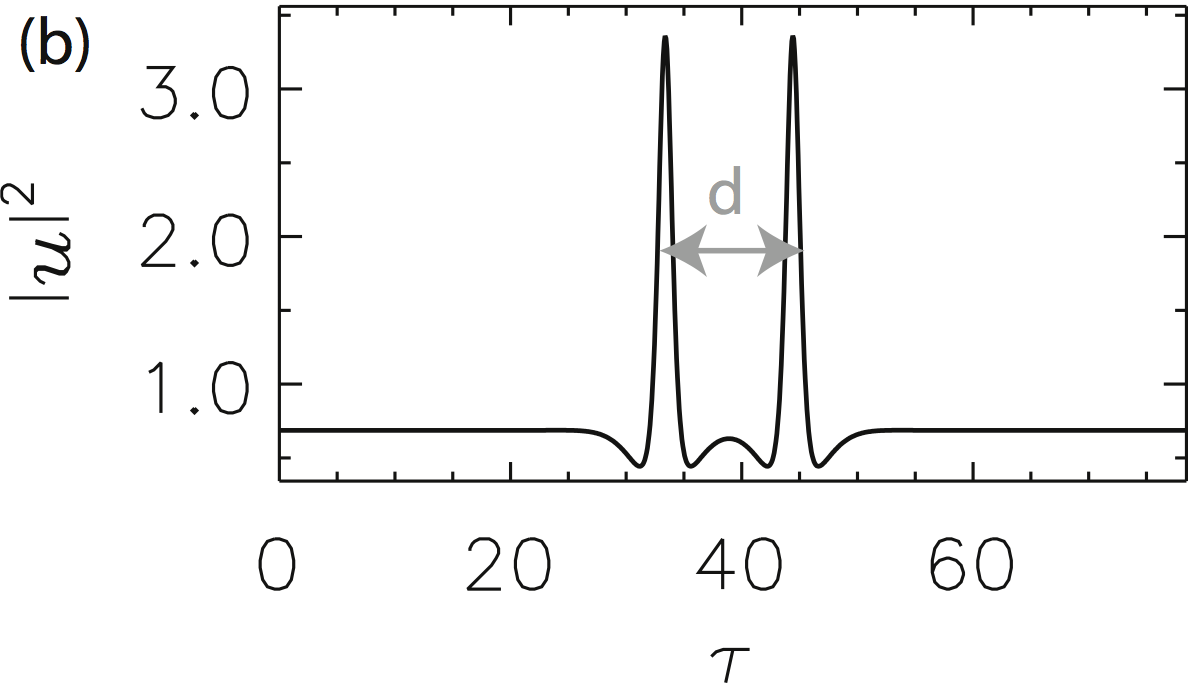}
\includegraphics[scale=0.5]{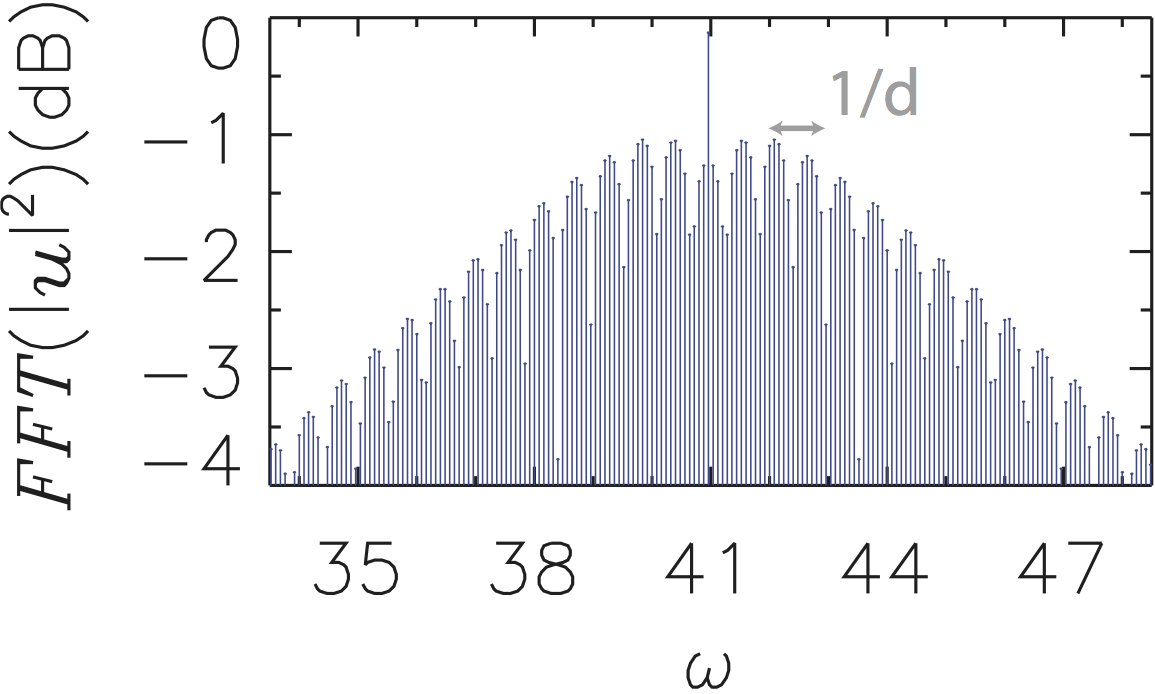}
\includegraphics[scale=0.5]{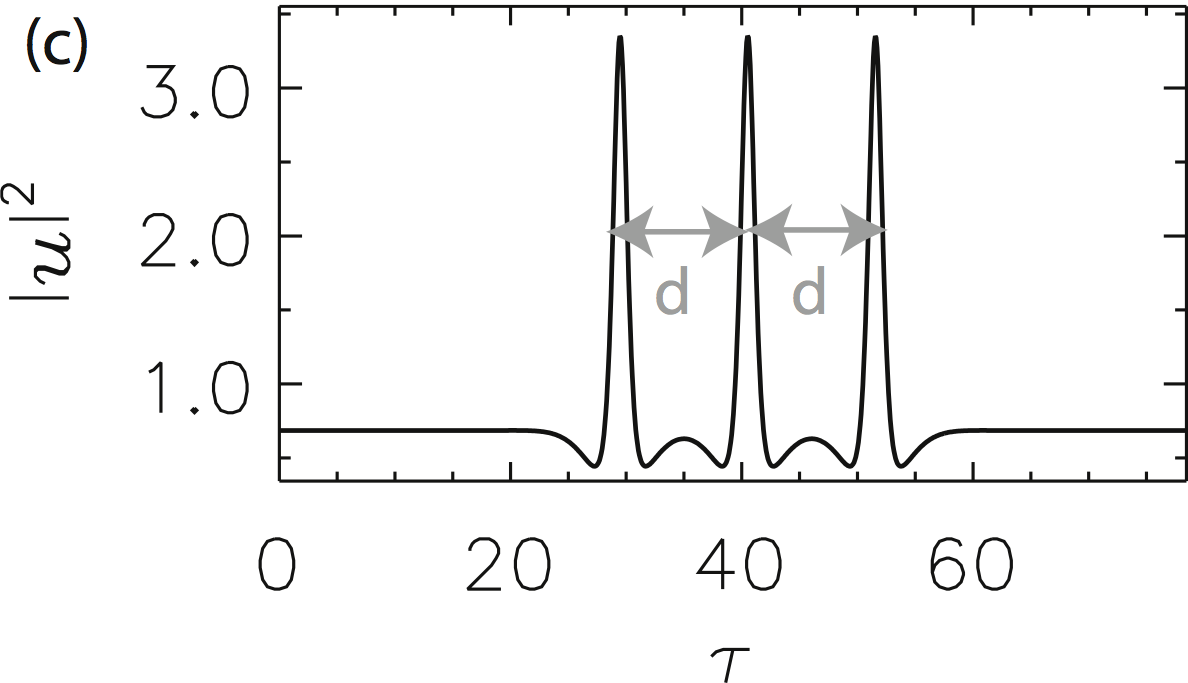}
\includegraphics[scale=0.5]{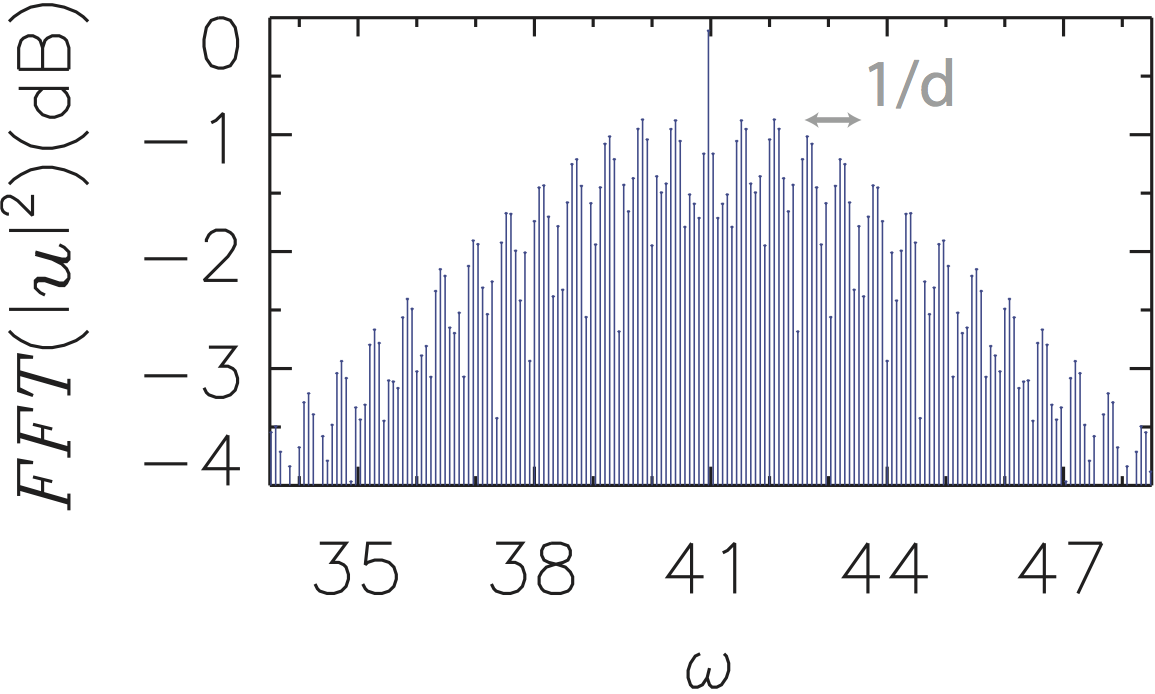}
\includegraphics[scale=0.5]{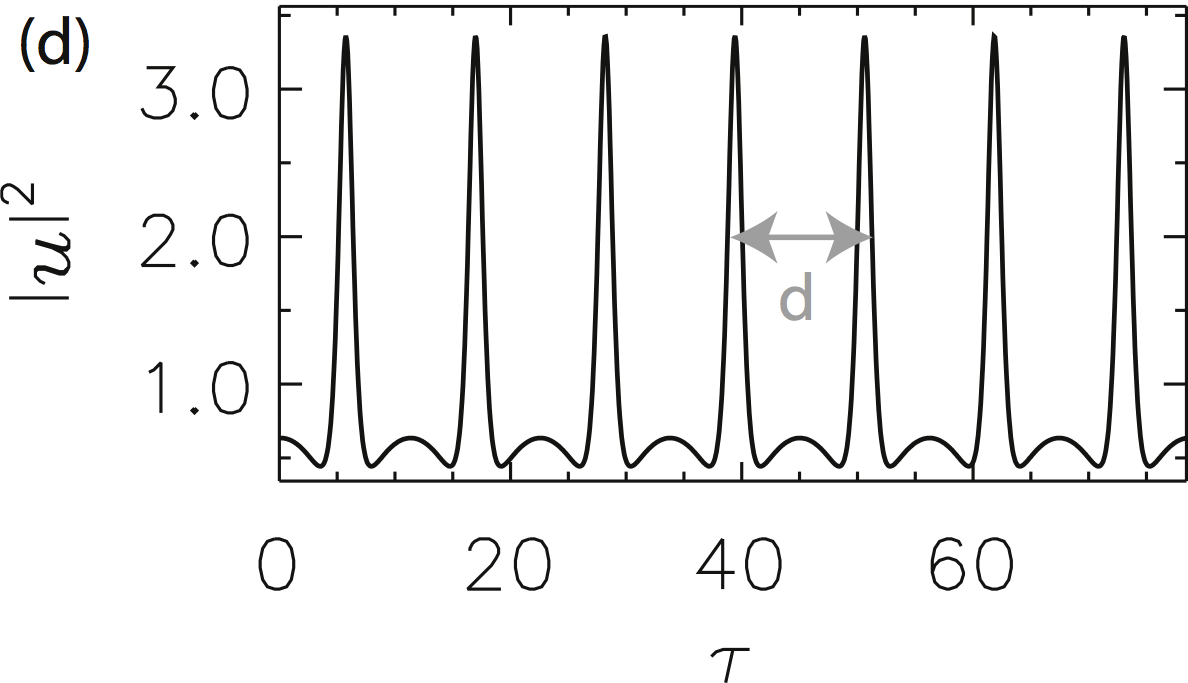}
\includegraphics[scale=0.5]{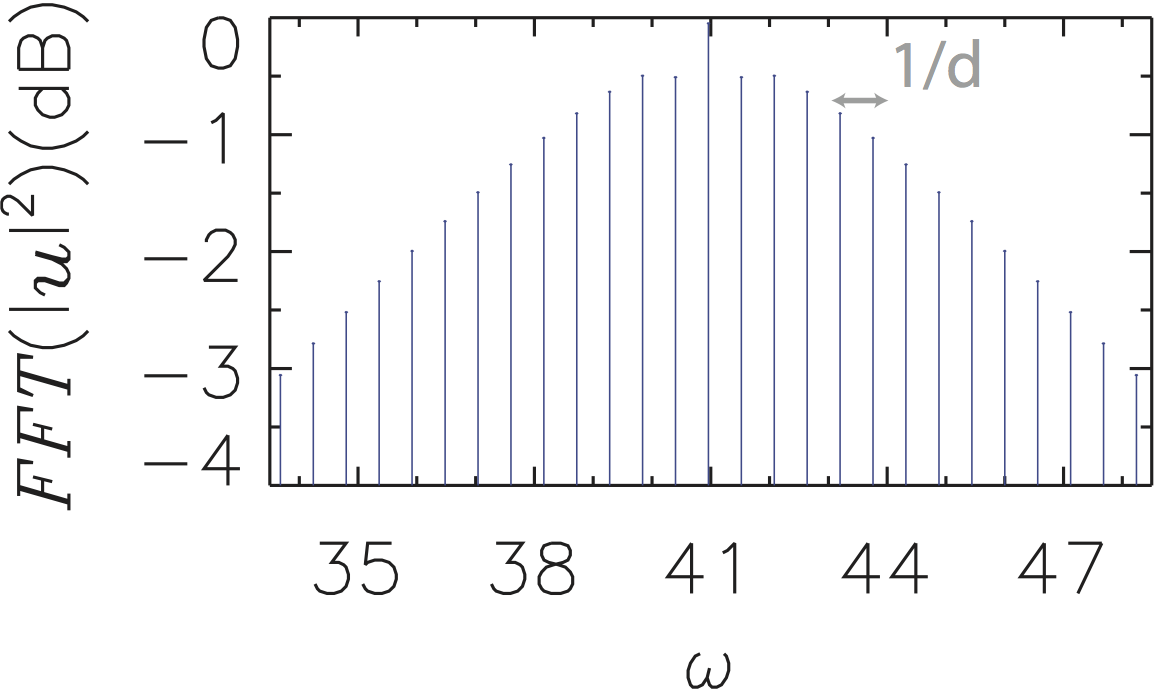}
\caption{Profiles of a single CS (a), two CSs (b), three CSs (c) and a pattern
(d) on the left hand side, with their corresponding frequency combs on the right
hand side. The system parameters are: $\theta = 1.8$, the free spectral range
$\mathrm{FSR} = 1/L = 0.08 $ and the number of discretization points was taken to be $N =
512$.}
\label{profiles}
\end{figure}

In Figure \ref{profiles}, we show typical profiles of the \ch{stable} CSs that exist in an optical cavity described
by the LLE. As mentioned before, there exist two branches of CSs that coexist in the pinning region. One branch
corresponds to \ch{CS states with an odd amount of peaks}, while the other branch contains the solutions with an
even amount of peaks. A typical one-peak CS is shown in Figure \ref{profiles}(a). In a fiber resonator or
microresonator this solution would correspond to a time-localized pulse circulating in the cavity. Such a light
pulse corresponds to a stable \ch{smooth} frequency comb in the corresponding frequency domain, as shown in the
panel on the right hand side. The distance between all frequency modes is given by the free-spectral-range
$\mathrm{FSR} = 1/L$, where $L$ is the cavity length, while the exact shape of the frequency comb is determined by
the Fourier transform of the profile of the CS itself. This equivalence has also been studied in Refs.\
\cite{coen_modeling_2012,coen_scalinglaws_2013,Chembo_LLE_2013}. Solutions with two peaks and three peaks are
plotted in panels \ch{(b) and (c)}, with the corresponding frequency combs. It can be seen that the effect of adding
extra peaks is to introduce an extra modulation of the frequency comb. \ch{The multiple peaks of CS bound states}
can only coexist at well-defined separation distances $d$ between them, determined by the typical wavelength of the
oscillatory tails of the \ch{CS peaks} \cite{Kozyreff_Gelens}. This separation distance $d$ therefore also
determines the modulation distance $1/d$ observed in the frequency comb. The modulation depth becomes more
pronounced as more and more peaks are added \ch{to the solution}.


In an infinite system (unbounded domain) CS peaks can be added indefinitely such that the two snaking branches in
principle continue indefinitely. However, realistic systems are always bounded. In this case the CS branches cannot
endlessly snake back and forth. Instead the snaking structure is truncated when the width of the
\ch{\emph{localized}} patterns consisting of multiple CSs approaches the domain size. In the case of fiber
resonators or microresonators, periodic boundary conditions apply and the system is finite as determined by the size
of the fiber loop, ring or disk resonator. In such a finite system the CS branches in the snaking region are
generally found to connect to a branch of patterned solutions. This is also the case here, as can be seen in Figure
\ref{fig::Hom_plus_snaking}. Moreover the branches of small amplitude CSs (i.e. the first unstable branch of single
CSs) no longer bifurcate from the HSS but now bifurcate in a secondary bifurcation on the primary branch of the
pattern solutions \cite{Knobloch_Bergeon_2008}. An example of a patterned solution with 7 peaks is shown in Figure
\ref{profiles}(d). All frequency modes present in the corresponding frequency comb are now separated by $1/d$ (or
equivalently 7 FSR units). Although we have used the same distance $d$ to denote the separation between various
peaks in the pattern, we remark that this should not necessarily be the case, as this might vary a bit depending on
the exact cavity length. We note also that CSs with $L/d-1$ peaks can be interpreted as one missing cell, or hole,
in a periodic pattern.

It is clear from Figure \ref{fig::Hom_plus_snaking} that nonlinear optical
cavities described by the LLE admit many stable and unstable localized and
patterned solutions for the same set of system parameters (such as the frequency
detuning and the pump power). Having such a highly multistable landscape has
important consequences if one aims at creating a stable Kerr frequency comb.
Every stable solution of the LLE has its own basin of attraction in the infinite dimensional phase space. Choosing the correct initial conditions and/or
writing process will thus be essential to target the structure one would like to
have in the nonlinear resonator. Minor changes might result in obtaining a
homogeneous solution, a single CS, a group of CSs or a pattern; all of which
correspond to different frequency combs.

In the next Section, we will analyze the spatial dynamics of the homogeneous
steady state solutions as this will provide valuable information about where one
can expect stable CSs to be found. We will verify this theory in Section
\ref{Sec::dyn1CS} and will examine the rich temporal dynamics of CSs, which is
also reflected in the corresponding frequency combs.

\section{Spatial dynamics}\label{Sec::spatdyn}


The solutions \ch{$u$} of the LLE in which we are interested, namely CSs and localized patterns (groups of CSs), are
stationary solutions of (\ref{eq.1}) (a PDE), i.e. they are such that $\partial_t u=0$ \footnote{This treatment
can also be extended for moving solutions, e.g. fronts, in the reference frame that travels with the solution.}.
Under this condition, Eq.~(\ref{eq.1}) can be simplified to a special dynamical 
system of first order real ODEs in which
space\ch{/fast-time}, $\varphi$ in our case, takes the role usually played by time (see, e.g. Refs.\
\cite{HaragusIoossbook,Champneys98,Gomila_PhysD,Colet_Gelens1}), namely,
\begin{eqnarray}
d_{\varphi} u_1&=&u_3 \nonumber  \\
d_{\varphi} u_2&=&u_4 \nonumber  \\
d_{\varphi} u_3&=& u_2+\theta u_1-u_1 I_s \nonumber \\
d_{\varphi} u_4&=&-u_1+\theta u_2-u_2 I_s+u_0\ ,
\label{eq:SpDyn}
\end{eqnarray}
with $u_1=\mathrm{Re}(u)$, $u_2=\mathrm{Im}(u)$, $u_3=d_{\varphi} u_1$, 
$u_4=d_{\varphi} u_2$, and $I_s=|u|^2=u_1^2+u_2^2$. Bounded trajectories of 
this dynamical system correspond to the stationary 
solutions of (\ref{eq.1}). In particular, it contains 
all information about the stability of the stationary solutions and describes 
what is known as their \emph{spatial dynamics} [spatial, because this 
technique has traditionally been applied to transverse structures]. The spatial dynamics describe also how different solutions can coexist (if at all) 
in the cavity and how each one connects to the other, hence revealing in which 
parts of the parameter space particular stationary solutions can potentially 
exist. Clearly, this has important practical implications for Kerr frequency 
combs.

The fixed points of Eqs.\ (\ref{eq:SpDyn}) are the HSSs $u_s$ of the original evolution equation (\ref{eq.1}), and 
their stability in space ${\varphi}$ can be analyzed by \ch{introducing} 
the ansatz $\ch{u} = u_s + \epsilon e^{\lambda {\varphi}}$
into Eqs.\  (\ref{eq:SpDyn}) \footnote{We recall that 
$\varphi$ can be either the transverse spatial coordinate, $x$, or the 
fast time, $\tau$, in the case of fibers and microresonators. In this 
article, we will keep referring to $\varphi$ as being a spatial coordinate to 
avoid confusion with the cavity time $t$.}. Keeping only linear terms 
in (small) $\epsilon$ leads to the
following condition for what are known as the spatial eigenvalues $\lambda$,
\begin{equation}\label{eq::spateig}
  \lambda^4+(4I_s-2\theta)\lambda^2+\theta^2-4\theta I_s +3I_s^2 +1=0\,.
\end{equation}

An important property of the spatial dynamics is 
reversibility, that stems from the invariance of the LLE
(\ref{eq.1}) with $\partial_t u=0$ under the ${\varphi} \rightarrow 
-{\varphi}$ transformation, and, equivalently, of
Eqs.\ (\ref{eq:SpDyn}) under the transformation $(u_1,u_2,u_3,u_4)\rightarrow (u_1,u_2,-u_3,-u_4)$  \cite{Gomila_PhysD}.
Note that this arises in the temporal Kerr comb problem ($\varphi=\tau$) 
because of our approximation to consider only
second order dispersion. Reversibility implies that the spatial eigenvalues 
always come in pairs (cf. e.g.
\cite{Colet_Gelens1})\ch{: each} spatial eigenvalue $\lambda$ is accompanied by its  counterpart $- \lambda$,
and this property manifests also in that Eq.\ (\ref{eq::spateig}) is only function of $\lambda^2$, not of $\lambda$ alone
\ch{[Eq.\ (\ref{eq::spateig}) is a biquadratic equation]} \footnote{By writing Eq.  
(\ref{eq:SpDyn}) in terms of a complex quantity it also appears clearly why the 
eigenvalue equation is biquadratic.}. Thus, a \ch{\emph{repelling}} eigenvalue 
is accompanied by another \emph{attracting} one with the same rate 
(repelling/attracting means that the spatial dynamics takes the
field of the solution away/towards the HSS $u_s$).

\begin{figure}[t!]
\centering
\includegraphics[scale=1]{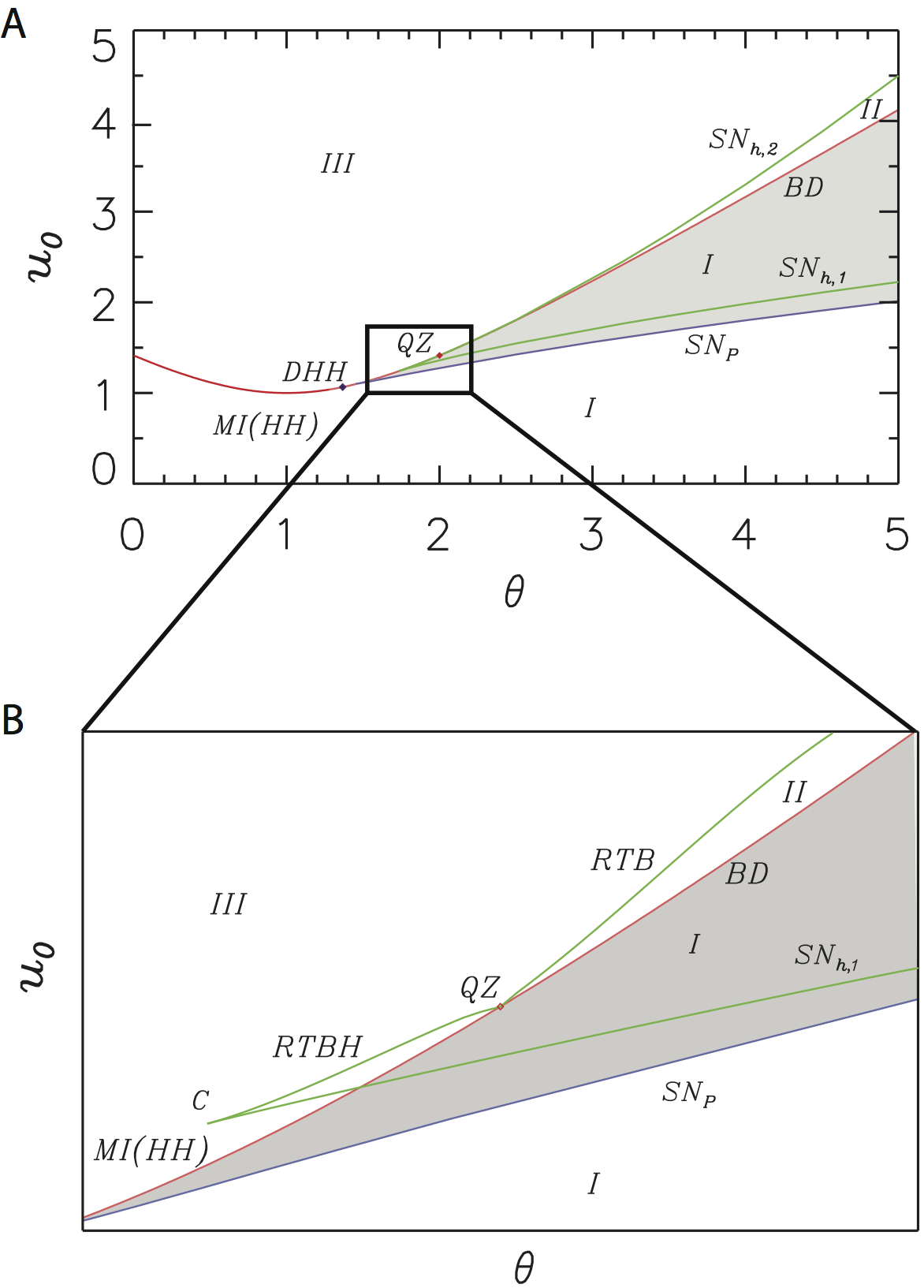}
\caption{(Color online) In Panel A all different regions that have a 
qualitatively different spatial eigenspectrum
are shown in the parameter plane ($\theta, u_0$), as obtained from Eq.\ (\ref{eq::spateig}). In the bistable region 
between $\mathrm{SN}_{h1}$ and
$\mathrm{SN}_{h2}$, \ch{$I_s$ in Eq.\ (\ref{eq::spateig}) is chosen such} that 
it corresponds to
the (stable) lower intensity HSS. The notation used is further clarified in 
Table \ref{tab:template}.
$\mathrm{SN}_P$ denotes the saddle-node bifurcation where the (stable) pattern 
is born. The light gray region
denotes the region where one can expect to find single and multiple CSs. Panel B shows a closer 
zoom 
of the organization of the spatial
eigenvalues around the Quadruple Zero (QZ) point where all eigenvalues are 
zero.} \label{fig::Spatial_dynamics}
\end{figure}

The spatial eigenvalues can be easily obtained by solving Eq.\ (\ref{eq::spateig}), that yields,
\begin{equation}\label{eq::spateigSolution}
 \lambda=\pm\sqrt{\theta-2I_s\pm\sqrt{I_s^2-1}}\ .
\end{equation}
These eigenvalues provide all relevant information about the HSSs, and allow to  
identify different regions in parameter space, in terms of the form of the leading 
eigenvalues, i.e. those whose real part is closer to zero. Three characteristic 
cases can be distinguished:

\begin{enumerate}

\item The eigenvalues are organized as a quartet of complex eigenvalues $\lambda 
= \pm q_0 \pm i k_0$, such that a trajectory leaving (and approaching) the HSS 
presents oscillations. This is the case corresponding to label I in Figure 
\ref{fig::Spatial_dynamics} and Table \ref{tab:template}.
\item The leading eigenvalues are a purely real doublet $\lambda = \pm q_0$, 
such that a trajectory leaving (and approaching) the HSS is monotonic (Label II 
in Figure \ref{fig::Spatial_dynamics} and Table 
\ref{tab:template}).
\item The leading eigenvalues are a purely imaginary doublet $\lambda = \pm i 
k_0$, such that the HSS is temporally unstable (Labels III and IV 
in Figure \ref{fig::Spatial_dynamics} and Table \ref{tab:template})
\end{enumerate}

\begin{table}[t]
\centering
\begin{tabular}{|l|c|c|c|c|}
\hline
Cod & $(\lambda_{1,2},\lambda_{3,4})$  &  Name  & Label  \\
\hline
Zero & $(\pm q_0\pm i k_0)$  & Double-Focus & I \\
\hline
Zero & $(\pm q_1,\pm q_2)$  & Double-Saddle & II \\
\hline
Zero & $(\pm ik_1,\pm ik_2)$  & Double-Center & III  \\
\hline
Zero & $(\pm q_0,\pm ik_0)$   & Saddle-Center & IV  \\
\hline
One & $(\pm q_0,0)$  & Rev. Takens-Bodganov & $\mathrm{SN}_{h,2}$(RTB)  
\\
\hline
One & $(\pm ik_0,0)$ & Rev. Takens-Bodgano-Hopf& $\mathrm{SN}_{h,2}$(RTBH)  \\
\hline
One & $(\pm q_0,\pm q_0)$  &  Belyakov-Devaney & BD \\
\hline
One & $(\pm ik_0,\pm ik_0)$  & Hamiltonian-Hopf & MI(HH)\\
\hline
Two & $(0,0)$   & Quadruple Zero& QZ \\
\hline
\end{tabular}
\caption{Labels of qualitatively different spatial eigenspectra and the
transitions between these regions.}
\label{tab:template}
\end{table}

Table \ref{tab:template} shows the various possibilities of how the eigenvalues can be organized and it also
contains the transitions between the different regions, which are codimension-$1$ and $2$ objects (indicated Cod in
the Table \footnote{Codimension refers to the number of conditions that have 
to be specified in parameter space, and, thus, codim-$1$ and -$2$ transitions 
are, respectively, lines and points in a $2$-dimensional parameter space.}.
For a more in-depth description of all transitions and their
names, we refer to Refs.\ \cite{Colet_Gelens1, Colet_Gelens2}. A similar analysis of the spatial dynamics
in the Lugiato-Lefever equation has recently also been reported in Ref.\ \cite{Chembo_arxiv}.

\begin{figure}[t!]
\centering
\includegraphics[scale=1]{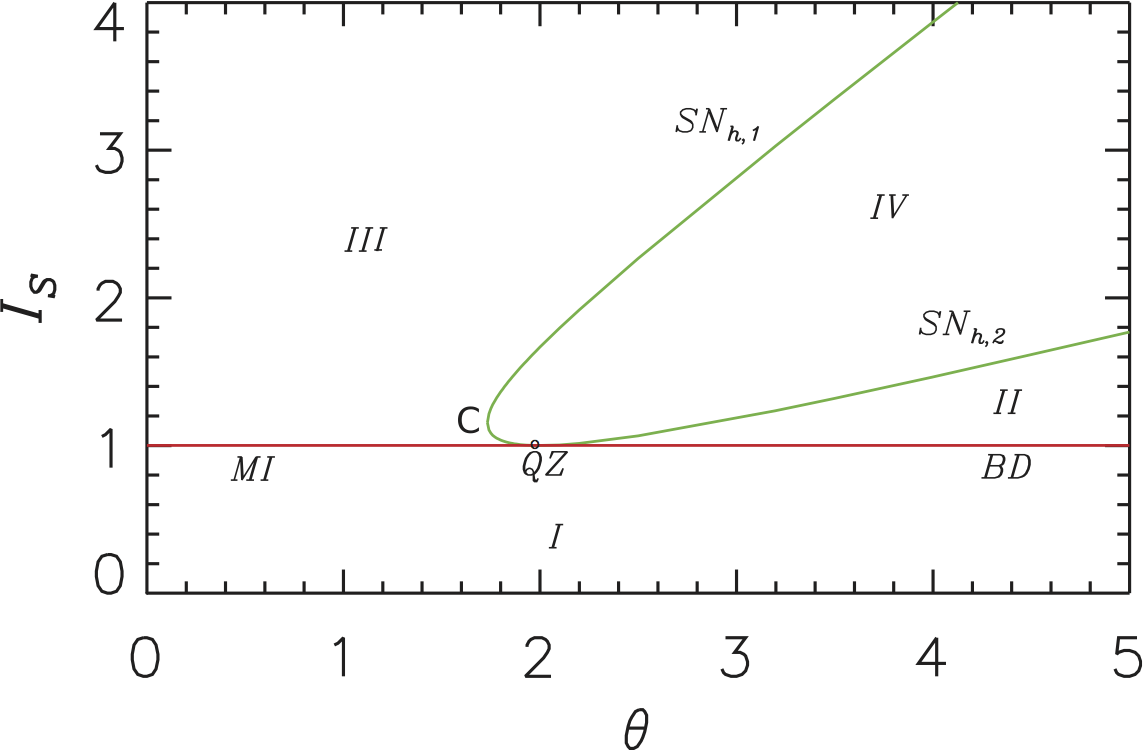}
\caption{(Color online) All different regions that have a 
qualitatively different spatial eigenspectrum
are shown in the parameter plane ($\theta, I_s$), as obtained from Eq.\ (\ref{eq::spateig}). The notation used is further clarified in 
Table \ref{tab:template}.} \label{fig::Spatial_dynamics_theta_IS}
\end{figure}

Figure \ref{fig::Spatial_dynamics_theta_IS} shows the organization of the spatial eigenvalues corresponding to the HSSs in the parameter plane ($\theta, I_s$).  In order to keep a closer connection to the experimental control parameters, however, we prefer plotting the structure of the spatial eigenvalues in the parameter plane ($\theta, u_0$), shown in Figure \ref{fig::Spatial_dynamics}. In the plane ($\theta, u_0$), we only report the spatial eigenvalues corresponding to the lower branch HSS in the bistable region for $\theta > \sqrt{3}$ \footnote{The upper branch HSS is always unstable in this case.}. In both figures, all labels are defined in Table \ref{tab:template}, allowing to understand the different regions. Here below, we will discuss the different regions in Figure \ref{fig::Spatial_dynamics} in more detail.

A pair of double zeros of Eq.\ (\ref{eq::spateig})  define 
a MI line for $\theta<2$  in parameter space (Figure \ref{fig::Spatial_dynamics}) 
given by
\begin{equation}\label{eq.22}
u_0=\sqrt{1+(1-\theta)^2}\ .
\end{equation}
This corresponds to $I_s=1$ (see also Figure \ref{fig::Spatial_dynamics_theta_IS}). 
This degenerate pair of double zeros is given by the purely imaginary eigenvalues $\lambda=\pm i\sqrt{2-\theta}$. In 
dynamical systems theory this bifurcation is known as a Hamiltonian-Hopf (HH). 
The Hamiltonian-Hopf bifurcation line separates the region I where the HSS is a 
double-focus (has a quartet of complex eigenvalues) from region III where the 
HSS is a double-center with four purely imaginary spatial eigenvalues. In this 
region III the HSS is unstable to spatiotemporal perturbations and patterned 
solutions develop.

The MI becomes a Belyakov-Devaney (BD) transition for $\theta>2$ through a  
Quadruple Zero (QZ) point at $\theta=2$, where $\lambda=0$ with multiplicity 
four. On the BD line the spatial eigenvalues are given by two pairs of
pure real eigenvalues $\lambda=\pm \sqrt{\theta-2}$. Note that, as the MI (HH), the BD also occurs for $I_s=1$, although, strictly speaking, it is not a real bifurcation 
as $\text{Re}(\lambda)\neq0$ at the transition. The BD line separates region I 
from region II, where the HSS is a double-saddle. After crossing this line into region II the HSS does 
not lose its stability and no patterned solutions develop until 
$\mathrm{SN}_{h,2}$ is crossed.

For $\theta>2$ the $\mathrm{SN}_{h,2}$ is a reversible Takens-Bogdanov (RTB) bifurcation with two zero spatial
eigenvalues and two real eigenvalues ($\pm q_0$). For $\theta<2$ the 
$\mathrm{SN}_{h,2}$ is a reversible Takens-Bogdanov-Hopf (RTBH). On the RTBH 
line two eigenvalues are zero and two are purely imaginary ($\pm ik_0)$ 
(see Figure \ref{fig::Spatial_dynamics}B and Table \ref{tab:template}).

\begin{figure}[t!]
\centering
\includegraphics[width = \columnwidth]{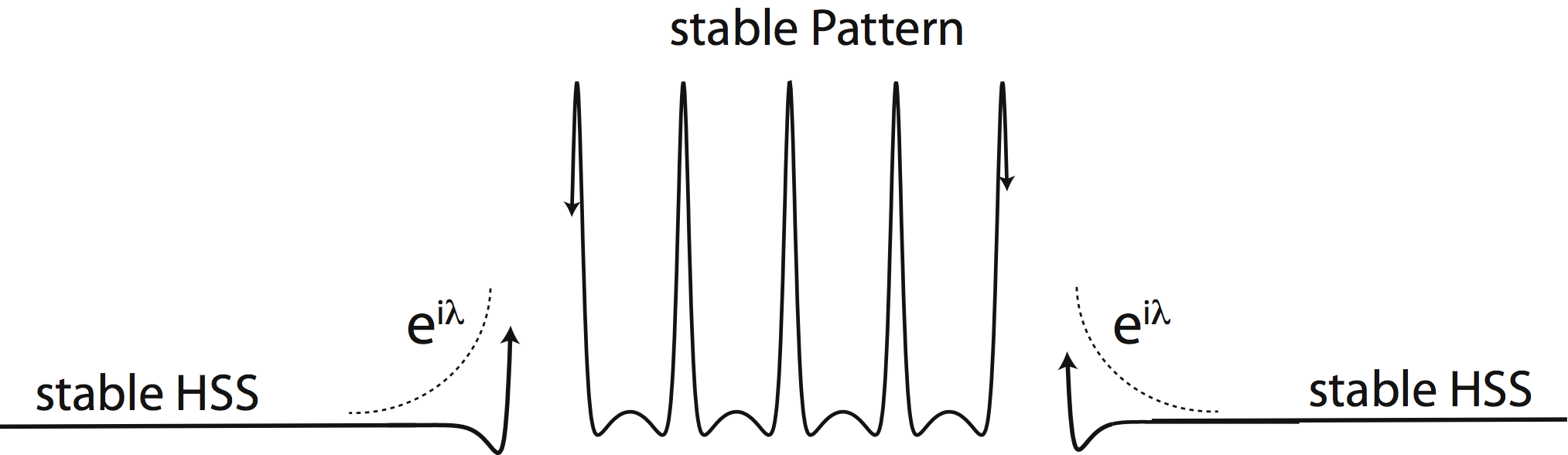}
\caption{Sketch of the spatial dynamics leading to CSs. A CS can be interpreted 
as a heteroclinic connection from a stable HSS to a stable pattern and another 
one from the pattern back to the HSS. The trajectory approaches the HSS 
according the leading spatial eigenvalues $\lambda$ of the HSS, which 
characterize the shape of the CS tails.}
\label{fig::Spatial_dynamics_sketch}
\end{figure}

In addition to the properties of HSSs, the spatial  eigenvalues also explain the asymptotic behavior of the CS profile as $\varphi \rightarrow \pm 
\infty$. The present work focuses on CSs that fall in the general class of 
localized structures that arise through the connection of a stable homogeneous 
state and a subcritical pattern \cite{Pomeau, Thual-1990, 
Tlidi_Taki2007,Woods, Coullet}. In the context
of the spatial dynamics, such localized solutions are homoclinic orbits to a 
fixed point (the HSS), that pass very close
(arbitrarily close increasing the number of peaks of the CS) to a periodic 
orbit (the patterned state), see Figure \ref{fig::Spatial_dynamics_sketch}. The 
shape of the front leaving and approaching the HSS (at the
same rate, due to reversibility) is given by the leading 
spatial eigenvalues found by solving (\ref{eq::spateig}). In order for 
(multiple) CSs to exist in any arbitrary order in the 
nonlinear resonator, it is required for the fronts
to have oscillatory tails, so that the tails avoid the merging and annihilation of CSs 
\cite{Colet_Gelens1,Colet_Gelens2}. Various CSs in the system are able to 
coexist in the cavity as they can lock to each other via
their overlapping oscillatory tails, and this has been shown to determine 
the possible locations (and separations) of CSs in a nonlinear cavity described 
by the LLE  \cite{Kozyreff_Gelens}. Eq.\ (\ref{eq::spateig}) should exhibit a 
complex quartet of eigenvalues in such a case.

\begin{figure}[t!]
\centering
\includegraphics[width = \columnwidth]{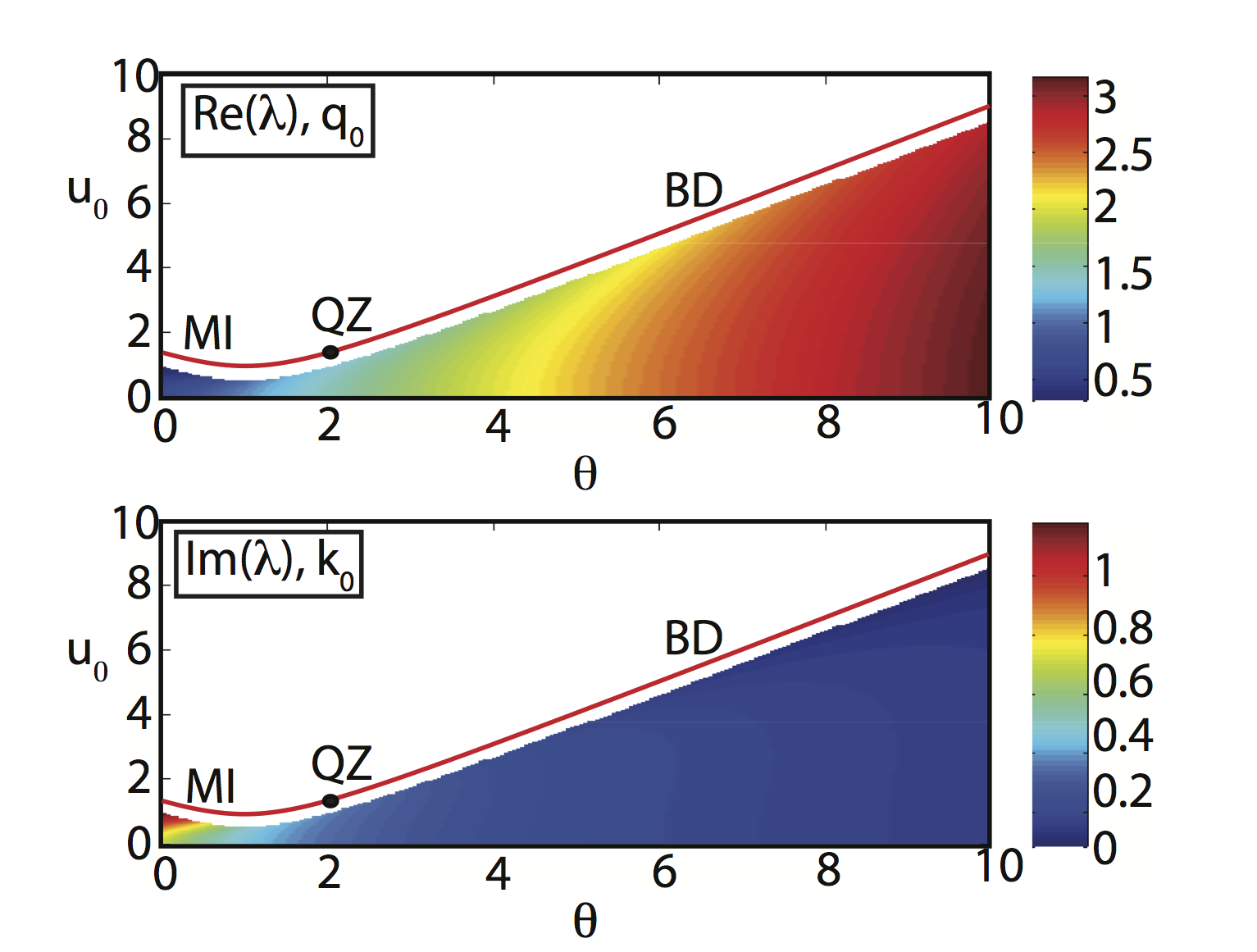}
\caption{(Color online) In the Region I, where all spatial eigenvalues are a quartet of complex eigenvalues $\lambda_s = \pm q_0 \pm i k_0$, we plot the real part $q_0$ and the imaginary part $k_0$ as a function of the detuning $\theta$ and the pump amplitude $u_0$.}
\label{fig::Spatial_eigenvalues}
\end{figure}

The region where the stable homogeneous solution has a
quartet of complex eigenvalues is  Region I. The 
gray region in Figure \ref{fig::Spatial_dynamics} is the
area where region I overlaps with the region of stable patterned solutions 
(above the $SN_p$ line \footnote{The $SN_p$ line in Figure 
\ref{fig::Spatial_dynamics} has been computed numerically, and it starts from 
the Degenerate Hamiltonian Hopf (DHH) point where the patterns become 
subcritical (see Section \ref{Sec::LLE}).}), such that all conditions are 
favorable to find CSs. Figure \ref{fig::Spatial_eigenvalues} provides more insight 
into how the real part $q_0$ and the imaginary part $k_0$ of the spatial 
eigenvalues of the stable HSS depend on the system parameters, $(\theta, u_0)$. One can see that in the 
bistable region\footnote{At $\theta = \sqrt{3}$, the system switches from being monostable to bistable. In Figure \ref{fig::Spatial_dynamics} this is shown by the Cusp bifurcation (denoted by the letter C) where the two saddle-node bifurcations of the HSSs ($SN_{h,1}$ and $SN_{h,2}$) are born.}, $\theta > \sqrt{3}$, $k_0$ is quite small and decreases as the 
detuning $\theta$ increases, while the real part $q_0$ strongly increases for 
higher $\theta$. Therefore, the oscillatory tails of a CS will be strongly 
damped at higher values of the detuning $\theta$.

Isolated localized structures, formed by either a single peak or a number of 
closely packed peaks, can also exist in Region II for intracavity intensities 
above $I_s=1$, since the stable HSS still coexists with patterned 
solutions. In this region, however, CSs have monotonic tails and two separated 
CSs will move towards each other and merge.

\section{Temporal dynamics of cavity solitons}\label{Sec::dyn1CS}

\begin{figure}[t!]
\centering
\includegraphics[width = \columnwidth]{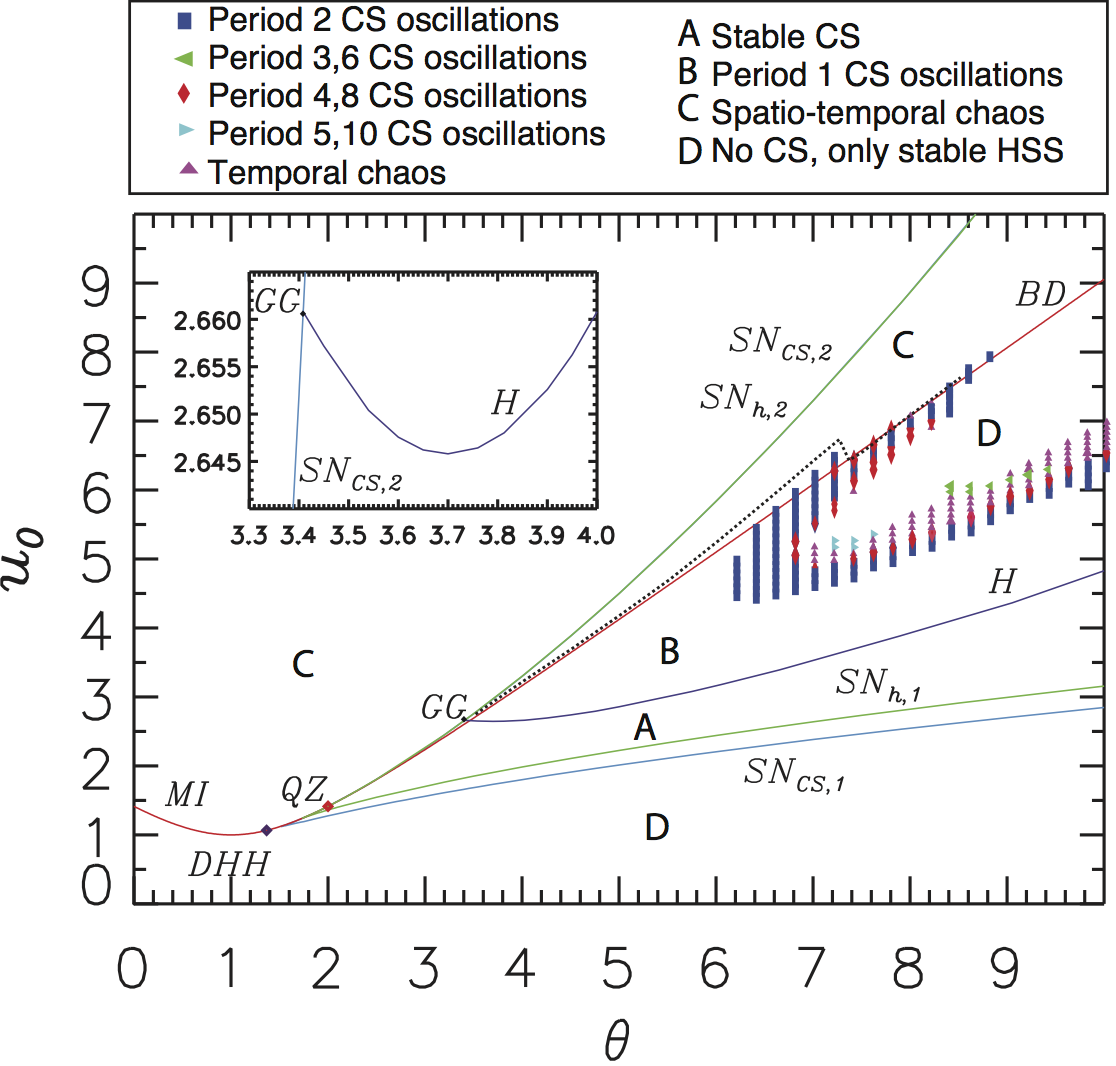}
\caption{(Color online) The temporal dynamics of single CSs are shown and
related to the transition lines obtained from the spatial dynamics also depicted
in Figure \ref{fig::Spatial_dynamics}. In region A single CSs are stable. Such
CSs can be destabilized by a Hopf bifurcation $H$ leading to time-oscillating
CSs, found in region B. In region C spatio-temporal chaos is found. In region D
only the stable HSS is found. Finally, more complicated time dynamics such as
temporal chaos is found in the region denoted by the markers. For more
information about the notation of the spatial dynamics, we refer to Table
\ref{tab:template}. The inset shows a zoom of the bifurcation lines around the
Gavrilov-Guckenheimer (GG) point.}
\label{fig::Temporal_dynamics}
\end{figure}

In this Section, we characterize the regions of existence of single CSs in 
the parameter space ($u_0,\theta$)
defined by the pump power and the cavity detuning. In Figure \ref{fig::Temporal_dynamics}, we have again plotted all
characteristic lines obtained from our analysis of the spatial dynamics in the previous Section (see also Figure
\ref{fig::Spatial_dynamics}). Moreover, by combining linear stability analysis and time evolution simulations, we have characterized not only the existence of CSs,
but also their temporal stability and dynamics (see also \cite{Leo_Gelens_2013}). Using a Newton-Raphson method we can numerically track the CS solutions in parameter space and calculate their temporal stability. This allows to accurately determine the location of the Hopf bifurcation line $H$. All other dynamical regimes have been determined numerically using time evolution simulations. In Region A, stable CSs can be
found. This region is delimited by the saddle-node bifurcation $\mathrm{SN}_{\mathrm{CS},1}$ (which largely
coincides with $\mathrm{SN}_{P}$ in the bistable region) and the Hopf bifurcation line $H$ where the CS is
destabilized. One can notice that the region of existence of stable CSs is largely within the 
gray region of Figure \ref{fig::Spatial_dynamics}.

\begin{figure}
\centering
\includegraphics[width = \columnwidth]{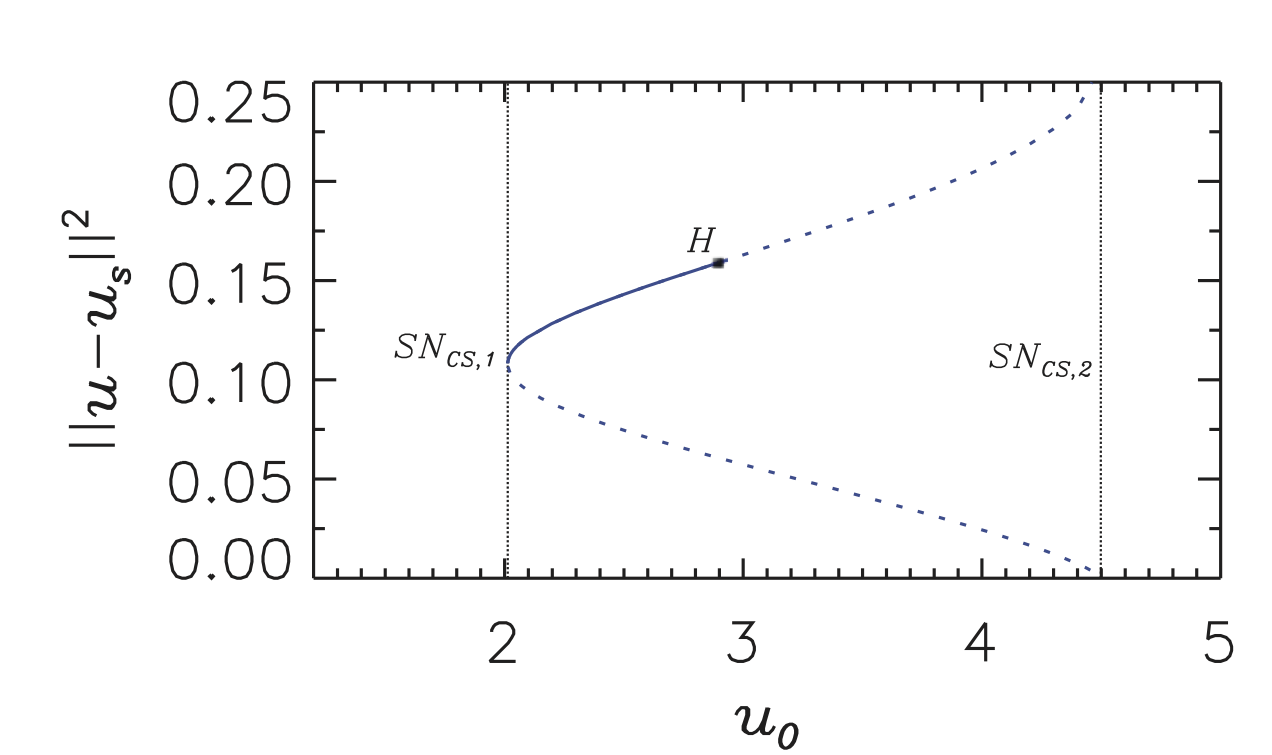}
\caption{Bifurcation diagram of the single peak
CS, showing the presence of a Hopf instability $H$ for $\theta = 5$. Solid
(dashed) lines represent stable (unstable) CSs.}
\label{fig::Hopf_bifdiagram}
\end{figure}

\begin{figure*}[t!]
\centering
\includegraphics[width = 2 \columnwidth]{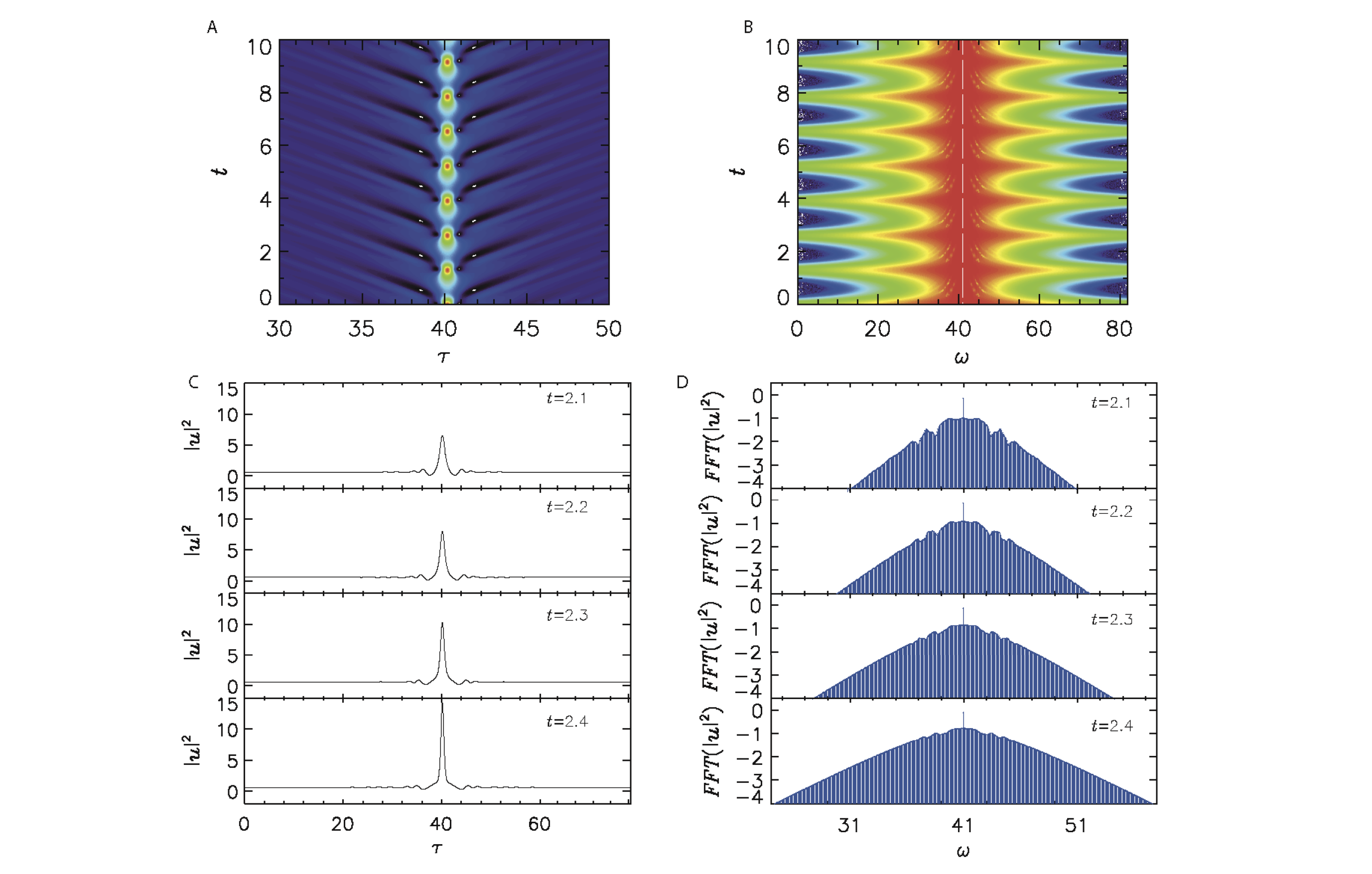}
\caption{(Color online) Spatio-temporal evolution of a CS and its
corresponding frequency comb is
plotted for $u_0 = 3.5$ and $\theta = 5$. Panels A and B show the time evolution in real and
frequency space respectively. Panels C and D show several profiles within one
oscillation period of the CS, both in the time (D) and frequency domain (E).}
\label{fig::Hopf_instability}
\end{figure*}

When crossing the Hopf bifurcation line $H$, the CS is no longer stable and displays time-periodic oscillations.
This region B has recently also been demonstrated experimentally \cite{Leo_Gelens_2013}. An example of such a Hopf
instability is shown in Figures \ref{fig::Hopf_bifdiagram} and \ref{fig::Hopf_instability}. Figure
\ref{fig::Hopf_bifdiagram} shows the bifurcation diagram of the single peak CS for a detuning $\theta  = 5$, showing
the presence of a Hopf instability $H$. Crossing this Hopf bifurcation, the CS still exists and remains localized,
but oscillates in time with a fixed period, see Figure \ref{fig::Hopf_instability}A. The corresponding frequency
comb thus also similarly oscillates in time as shown in Figure \ref{fig::Hopf_instability}B. Examples of how both
the CS profile and the frequency comb change during one oscillation period are given in Figure
\ref{fig::Hopf_instability}C-D. As the CS oscillates, the envelope of the frequency comb is modulated. Such
modulation becomes more pronounced as the amplitude of the CS oscillation increases with increasing values of the
pump power $u_0$ and detuning $\theta$. The simultaneous occurrence of a Hopf bifurcation and a saddle-node
bifurcation, determined by the three temporal eigenvalues $\lambda_{a,b}=\pm i\omega$ and $\lambda_c=0$, with
$\omega>0$, defines the codimension-$2$ point known as Gavrilov-Guckenheimer (GG) or Fold-Hopf bifurcation  \cite{Guckenheimer_2002}. Thus, the Hopf
bifurcation line emerges from the GG point, and although the Hopf line looks like it terminates perpendicularly to
$\mathrm{SN}_{\mathrm{CS},2}$, the inset in Figure \ref{fig::Temporal_dynamics} shows that it in fact approaches the
GG point in a tangential manner.

\begin{figure*}[t!]
\centering
\includegraphics[width = 2  \columnwidth]{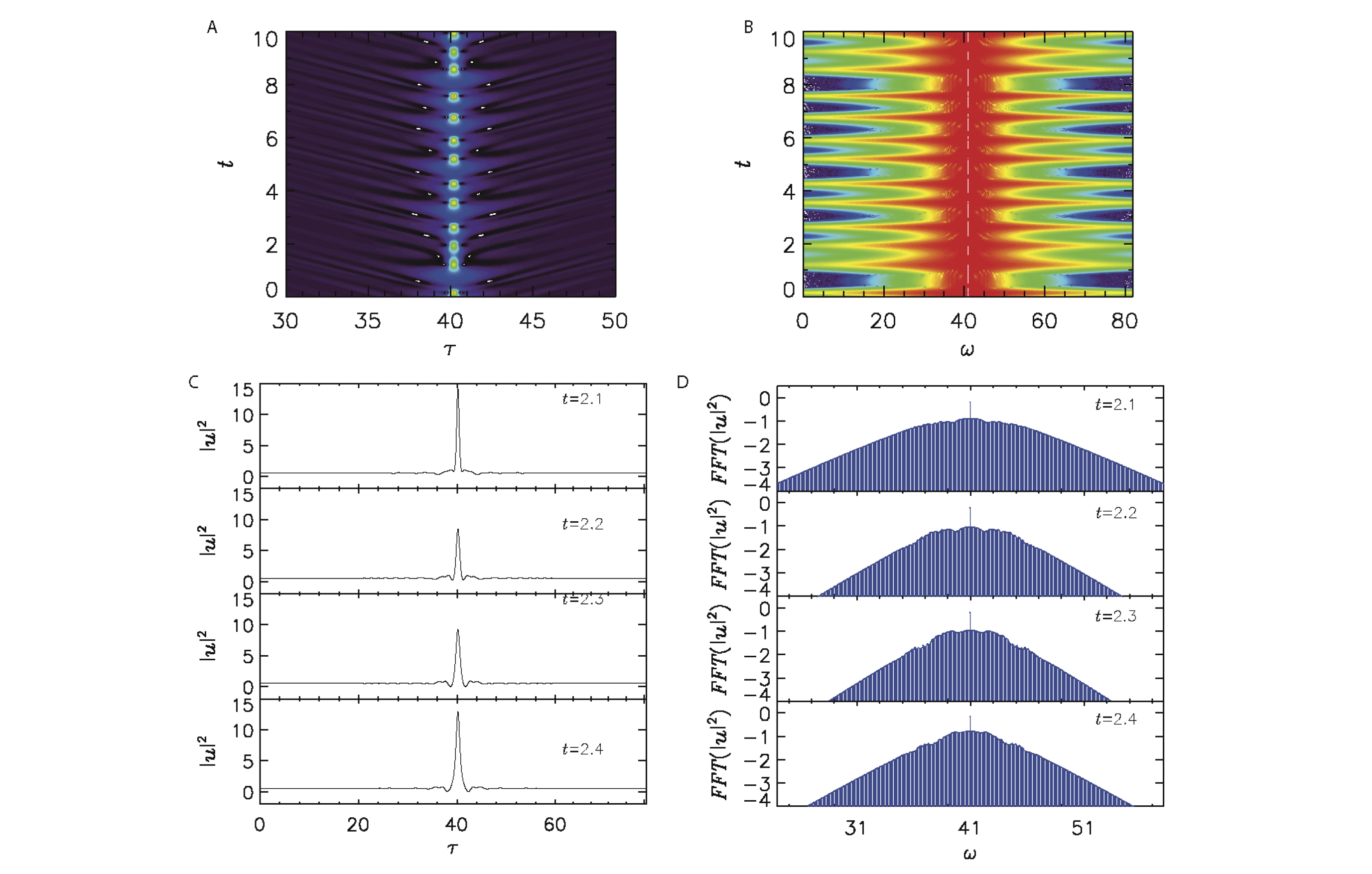}
\caption{(Color online) Spatio-temporal evolution of a CS and its
corresponding frequency comb is
plotted for $u_0 = 6.9$ and $\theta = 10$. The panels show the same as in Figure \ref{fig::Hopf_instability}, but now for
temporal chaos.}
\label{fig::chaos}
\end{figure*}

\begin{figure*}[t!]
\centering
\includegraphics[width = 2  \columnwidth]{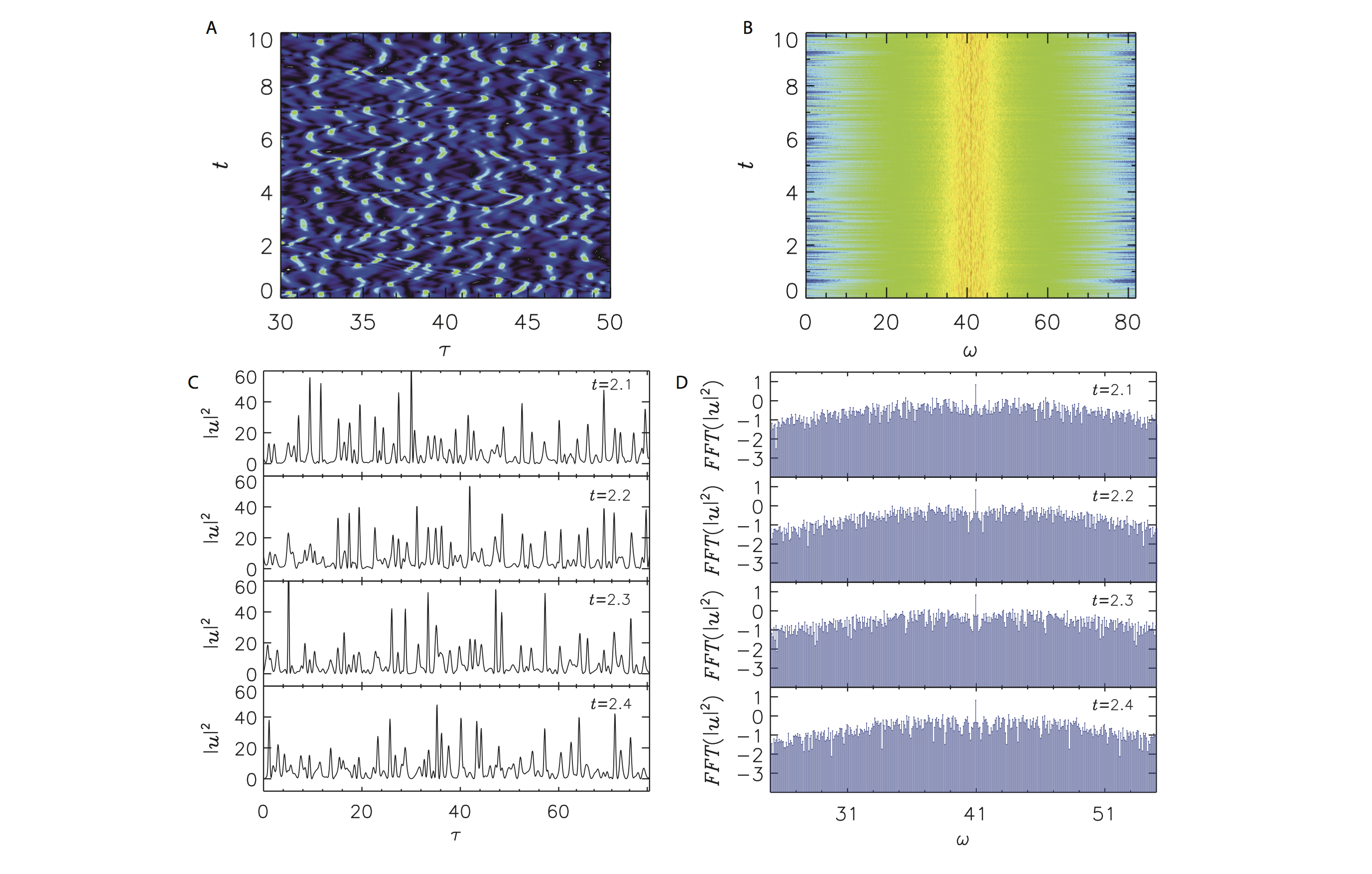}
\caption{(Color online) Spatio-temporal evolution of a CS and its
corresponding frequency comb is
plotted for $u_0 = 8$ and $\theta = 8$. The panels show the same as in Figure \ref{fig::Hopf_instability}, but now for
spatio-temporal chaos.}
\label{fig::chaos2}
\end{figure*}

For higher values of the detuning more complex oscillatory behavior can be found as well, such as period-2
oscillations, period-3 oscillations, period-N oscillations and eventually temporal chaos \cite{Leo_Gelens_2013}. An
example of such temporal chaos is given in Figure \ref{fig::chaos} where one can see that the CS remains localized,
but its envelope is oscillating in a chaotic way in time. Likewise, the envelope of the corresponding frequency comb
will change chaotically in time. For higher values of the pump power $u_0$ one also encounters large regions with a
different type of chaotic response, called spatio-temporal chaos or optical turbulence, as denoted by the letter C
in Figure \ref{fig::Temporal_dynamics}. An example of such chaotic behavior is shown in Figure \ref{fig::chaos2},
where it is clear that one can no longer distinguish clear localized CSs. Instead the entire cavity is oscillating
in a chaotic manner. The envelope of the corresponding frequency comb is no longer smooth but has significant
variations of the amplitude from mode to mode due to the many different wavelengths participating in the dynamics,
and it is much flatter. This regime is the one dominating the parameter space for large input powers. It can be
reached directly from the HSS crossing the line $\mathrm{SN}_{h,2}$ (Figure \ref{fig::Temporal_dynamics}) or through
the instabilities of a CS for powers around the $BD$ line. The spatiotemporal chaotic regime coexists with the HSS
for a range of input powers below $\mathrm{SN}_{h,2}$. The exact scenario that leads
to such spatio-temporal chaos coming from a CS needs further characterization, but is seems \ch{to be} closely
related to the Belyakov-Devaney (BD) line where the oscillatory feature 
associated to the spatial eigenvalues of the HSS is
lost and the tails of the CS approach the HSS in 
a monotonic manner. The origin of the temporal chaos observed for
higher values of detuning is also currently unknown, but might originate from the unfolding of the GG point
\cite{Gaspard_chaos}. These two points will be studied in more detail elsewhere.

\section{Concluding remarks}\label{Sec::discussion}

We have presented a comprehensive overview of the dynamics of the
LLE with one transverse dimension in relation with the generation of frequency
combs. Many different dynamical regimes, going from a single peak stationary CS
to spatio-temporal chaos, are available by changing the input pump and detuning
as summarized in Figure \ref{fig::Temporal_dynamics}.
On the one hand the LLE finds applications to many different
nonlinear optical cavities, and on the other hand it is a prototypical model for the
study of cavity solitons. Therefore many of the conclusions concerning the
generation of frequency combs from the different dynamical regimes are
applicable to other systems displaying CSs.

Due to the fact that many different dynamical regimes are supported in the LLE, this model offers several different
possibilities for the generation of frequency combs. In particular the coexistence of multiple solutions differing
only in the number of peaks circulating in the cavity offers the possibility to tune the shape of the comb without
changing the parameters of the system. To target particular solutions, suitable initial conditions must be seeded in
the cavity though. \ch{This observation may explain the variations in the shape of Kerr frequency combs observed in
experiments with microresonators after the pump field is interrupted and restarted.}

Beyond regimes of stationary CSs,  in very wide parameter range the LLE displays dynamical regimes, going from
regularly oscillating CSs to chaotic CSs and spatiotemporal chaos. Since these regimes appear for higher values of
the input pump amplitude and the detuning, they have a much broader bandwidth. The amplitude of the Fourier modes
is, however, not constant \ch{over time. Given the size of the parameter space associated with these solutions, one
cannot preclude that some reported experimental spectra have actually been obtained in this regime and, owing to the
slow speed of optical spectrum analyzers, are therefore averaged over the dynamics.}\\


\section{Acknowledgments}

We thank P. Colet and F. Leo for stimulating discussions. This research was supported by the Research Foundation - Flanders
(FWO), by the Spanish MINECO and FEDER under Grants FISICOS (FIS2007-60327) and
INTENSE@COSYP (FIS2012-30634),  by Comunitat Autonoma de les Illes Balears, and by the Belgian Science Policy Office (BelSPO) under Grant No. IAP 7-35. S.~Coen also acknowledges the support of the
Marsden Fund of the Royal Society of New Zealand.

%

\end{document}